\documentclass[%
reprint,
 amsmath,amssymb,
 aps,
]{revtex4-1}

\usepackage{graphicx}
\usepackage{dcolumn}
\usepackage{bm}

\begin{document}

\preprint{APS/123-QED}

\title{
Thermal Hall conductivity and topological transition in a chiral $p$-wave superconductor for Sr$_2$RuO$_4$ }%

\author{Yoshiki Imai$^{1}$}%
\email{imai@phy.saitama-u.ac.jp}
\author{Katsunori Wakabayashi$^{2,3}$}%
\author{Manfred Sigrist$^{4}$}
\affiliation{$^{1}$Department of Physics, Saitama University, Saitama 338-8570, Japan}
\affiliation{$^{2}$International Center for Materials Nanoarchitectonics (WPI-MANA), National Institute for Materials Science (NIMS), Tsukuba 305-0044, Japan}
\affiliation{$^{3}$Kwansei Gakuin University, Sanda 669-1337, Japan}
\affiliation{$^{4}$Theoretische Physik, ETH-Z\"urich, CH-8093 Z\"urich, Switzerland}

\date{\today}

\begin{abstract}
The interplay between the thermal transport property and the topological aspect is investigated in a spin-triplet chiral $p$-wave superconductor Sr$_2$RuO$_4$ with the strong two-dimensionality. 
We show the thermal Hall conductivity is well described by the temperature linear term and the exponential term in the low temperature region. 
While the former term is proportional to the so-called Chern number directly, the latter is associated with the superconducting gap amplitude of the $\gamma$ band.  
We also demonstrate that the coefficient of the exponential term  changes the sign around Lifshitz transition. 
Our obtained result may enable us access easily the physical quantities and the topological property of Sr$_2$RuO$_4$ in detail. 
\begin{description}
\item[PACS numbers]
\end{description}
\end{abstract}

\pacs{Valid PACS appear here}
\maketitle

\section{Introduction}
The transition metal oxide superconductor Sr$_2$RuO$_4$~\cite{maeno94,mackenzie03,maeno12} has attracted much interest as a strong candidate for topological superconductivity. 
The superconducting state is characterized by spin-triplet Cooper pairing~\cite{ishida98} and broken time-reversal symmetry~\cite{luke98,xia06}. 
Within the traditional symmetry classification scheme the order parameter with chiral $p$-wave Cooper pairing remains as the only phase compatible with these experiments: 
$
\bm d (\bm k)= \Delta_0 \hat{z}(k_x\pm ik_y) 
$
characterized by an orbital angular momentum $L_z = \pm 1$ along the $z$-axis \cite{rice95}. This is an analog to the A-phase of superfluid $^3$He which in a quasi-two-dimensional system as Sr$_2$RuO$_4$  opens a full quasiparticle excitation gap and whose topological character can be labeled by Chern number.

Angle-resolved photoemission-spectroscopy (ARPES) and de Haas-van Alphen data~\cite{damascelli00,mackenzie96,bergemann00} as well as first principles calculations reveal that the $\gamma$ band, a genuinely cylindrical (two-dimensional) electron-like Fermi surface and derived from the Ru $4d$-$t_{2g}$ $d_{xy}$ orbital, has its Fermi level only slightly below the van Hove singularity.  This means that the Fermi surface approaches the Brillouin zone boundary closely at the saddle points ($\pi/a,0$) and ($0, \pi/a$) with a lattice constant $a$, such that Sr$_2$RuO$_4$ could be close to a Lifshitz transition. 
In the superconducting phase the energy gap is suppressed strongly by symmetry close to these van Hove points. 
Topology induced features, therefore, are fragile against disorder effects and thermal broadening which destroy the quasiparticle gap~\cite{wang13,imai13}. 

At $c$-axis oriented surfaces of Sr$_2$RuO$_4$ a lattice reconstruction occurs, whereby the RuO$_6$ octahedra rotate around their $c$-axis leading to the doubling of the unit cell~\cite{Matzdorf00,veenstra13}. 
We have pointed out recently that such a rotation may change the Fermi surface topology of the $\gamma$ band pushing the Fermi energy through the van Hove points and in this way generate a Lifshitz transition between an electron- and hole-like shape~\cite{imai14}. 
In addition, the studies of the electron doping effect of Sr$_2$RuO$_4$ by La substitution for Sr ions~\cite{kikugawa04A,kikugawa04B,shen07}, the uniaxial pressure and the strain effects on Sr$_2$RuO$_4$~\cite{kittaka10,hicks14,taniguchi15} show that the topology of the $\gamma$ band changes around the van Hove points in the whole bulk system. 
In the chiral superconducting phase topological property depends on the structure of the $\gamma$ band~\cite{imai13}, and the different Fermi surface topologies give different Chern numbers.

It has been shown that the chiral $p$-wave superconductor supports chiral edge states responsible for spontaneous in-plane supercurrents along the surfaces~\cite{matsumoto99,furusaki01}. These current, however, are not a direct feature of topology, in the sense that their magnitude depends strongly on the band structure and the orientation and quality of the surface \cite{imai14,bouhon14,huang15}.  It has been shown that surface currents are essentially insensitive to changes of Chern number, e.g. through a Lifshitz transition~\cite{imai14}. Moreover, surface currents need not flow in a simple circular pattern around a disk-shaped sample as naively expected, but can even show peculiar current reversals \cite{bouhon14}. Thus, surface supercurrents are not an optimal feature to test the topology of the superconducting phase. 

On the other hand, the thermal Hall effect (Righi-Leduc effect) is more suitable to study the topology of the superconducting phase, as it realizes quantization features which are directly connected with the Chern numbers. 
In the following we will discuss the thermal Hall effect in the context of topology, using a lattice fermion model with an attractive inter-site interaction within the BCS-type mean-field approximation. 
Moreover, we analyze the temperature-dependence of the thermal Hall conductivity, in particular, near the Lifshitz transition changing the Chern number.

\section{Model }

The electronic properties of Sr$_2$RuO$_4$ are governed by the Ru 4$d$-$t_{2g}$ orbitals, and the electron bands yield the three Fermi surfaces, $ \alpha $, $ \beta $ and $ \gamma $. 
The hole-like $\alpha$ and the electron-like $\beta$ bands have one-dimensional characters, whose Fermi surface topologies are robust under a change of the chemical potential and topological properties are opposite to each other and, thus, vanish~\cite{raghu10,imai12}. 
Therefore, for simplicity, we will focus here on the $ \gamma $ band which remains the only one essential for topological aspects. 

\begin{figure}[b]
\begin{center}
\includegraphics[height=40mm]{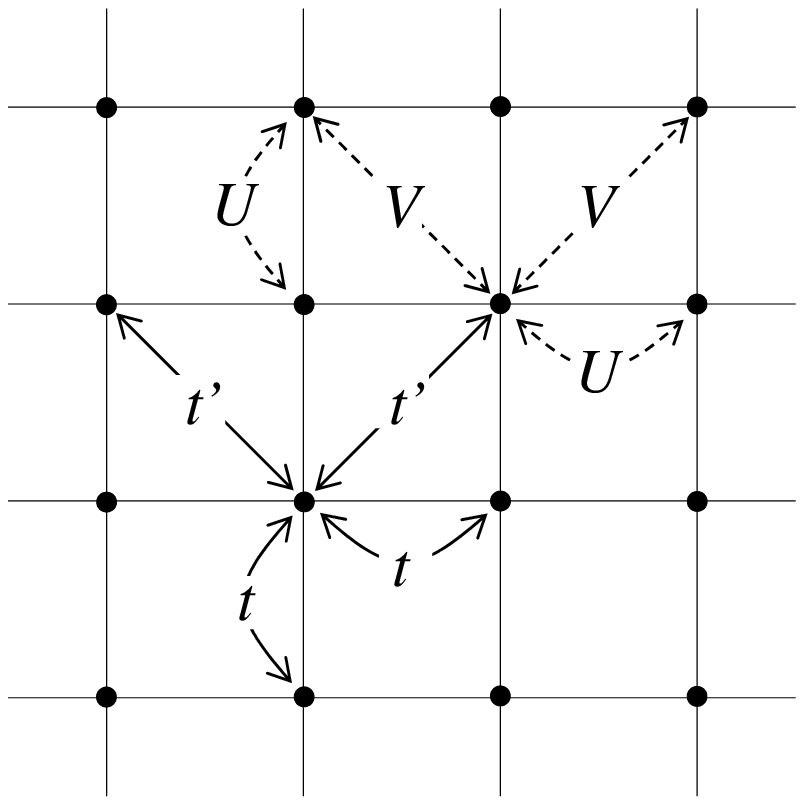}
\includegraphics[height=35mm]{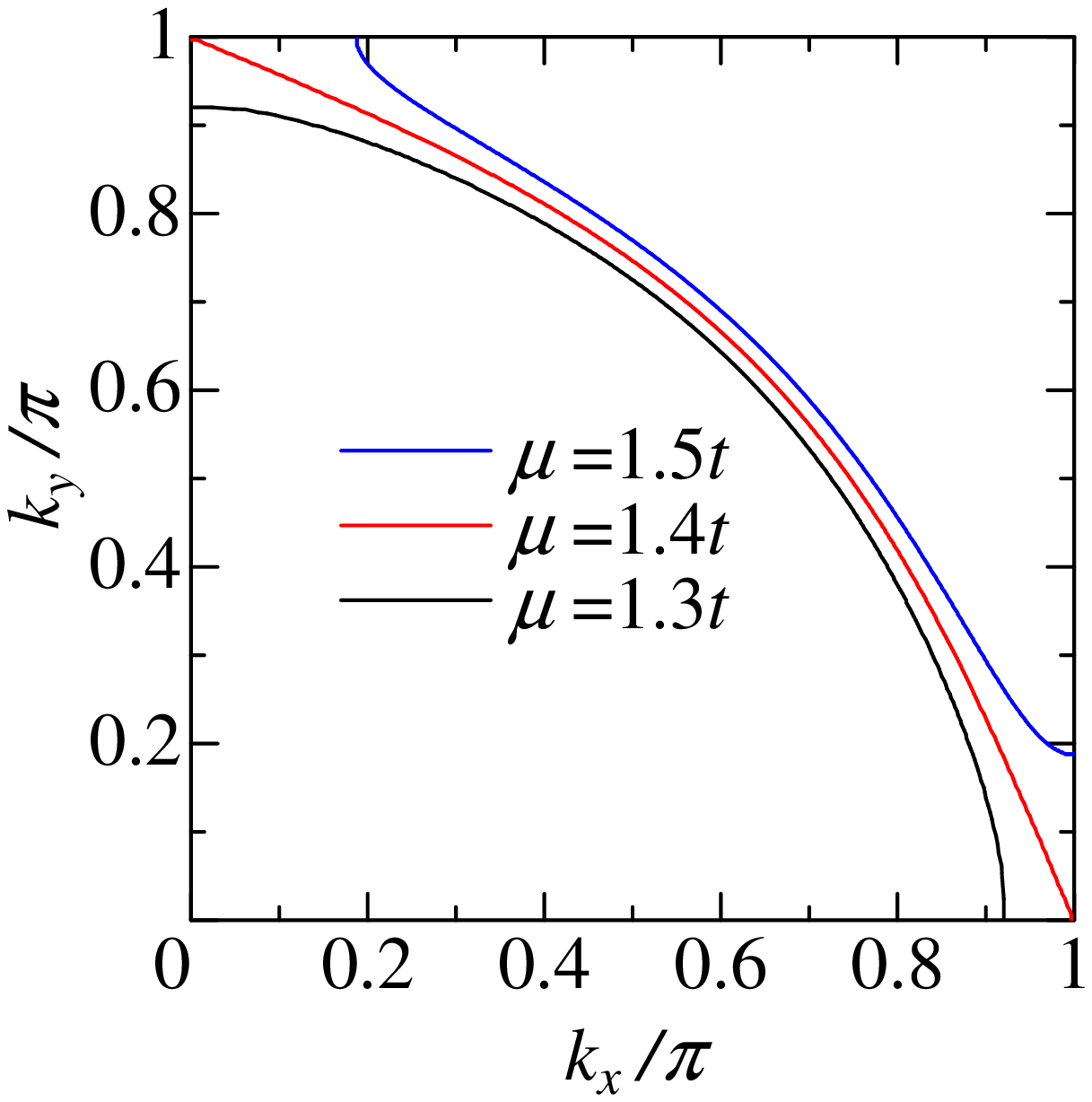}
\caption{(Color online) (Left panel) Lattice structure with square lattice. $t$ ($t'$) stands for the hopping amplitude between nearest (next nearest) neighbor lattice sites. 
$U$ ($V$) represents the attractive interaction between nearest (next nearest) neighbor lattice sites. (Right panel) The Fermi surface for several choices of the chemical potential $\mu$ in the normal phase. }
\label{lattice}
\end{center}
\end{figure}
The left panel of Fig. \ref{lattice} shows the lattice structure used for our model, whereby $t$ ($t'$) and $U$ ($V$) denote the hopping amplitudes and the coupling constants of the attractive interaction between nearest neighbor (next nearest neighbor) sites. These two parameters provide an enlarged range of topological superconducting states with different Chern numbers.  

For this single-band tight-binding model, we introduce the following Bogoliubov-de Gennes (BdG) Hamiltonian yielding a spin-triplet pairing on a two-dimensional square lattice 
\begin{eqnarray}
H^{\rm BdG}=\sum_{\bm k}
\begin{pmatrix}
c^{\dag}_{\bm k\uparrow}, c_{-\bm k\downarrow}
\end{pmatrix}
\begin{pmatrix}
\varepsilon_{\bm k} & \Delta_{\bm k} \\
\Delta_{\bm k}^* & -\varepsilon_{\bm k}
\end{pmatrix}
\begin{pmatrix}
c_{\bm k\uparrow}\\
c^{\dag}_{-\bm k\downarrow}
\end{pmatrix}
\label{eqn:BdG}
,
\end{eqnarray}
where $c^{\dag}_{\bm k\sigma}$ ($c_{\bm k\sigma}$) is the creation (annihilation) operator of an electron with wave vector $\bm k$ ($=(k_x,k_y)$) and spin $\sigma (=\uparrow, \downarrow)$. The  normal state electron dispersion of the $\gamma$ band and the gap function are parametrized as  
\begin{eqnarray}
\varepsilon_{\bm k}&=&-2t(\cos k_x +\cos k_y) -4t' \cos k_x \cos k_y-\mu, \\
\Delta_{\bm k}&=& 2iU \left\{\Delta_x \sin k_x +\Delta_y\sin k_y\right\}\nonumber \\
&+&2iV\left\{\Delta_{+}\sin (k_x+k_y) + \Delta_{-}\sin (-k_x+k_y)\right\}, 
\end{eqnarray}
%
where $\mu$ is the chemical potential.
Moreover, the symbols 
$\Delta_{m(=x,y,+,-)}$ represent the spin-triplet superconducting order parameter as follows, 
\begin{eqnarray}
\Delta_x&=&\frac{1}{2}(\langle c_{i\uparrow}c_{i_x\downarrow}\rangle+\langle c_{i\downarrow}c_{i_x\uparrow}\rangle),\\
\Delta_y&=&\frac{1}{2}(\langle c_{i\uparrow}c_{i_y\downarrow}\rangle+\langle c_{i\downarrow}c_{i_y\uparrow}\rangle),\\
\Delta_+&=&\frac{1}{2}(\langle c_{i\uparrow}c_{i_+\downarrow}\rangle+\langle c_{i\downarrow}c_{i_+\uparrow}\rangle),\\
\Delta_-&=&\frac{1}{2}(\langle c_{i\uparrow}c_{i_-\downarrow}\rangle+\langle c_{i\downarrow}c_{i_-\uparrow}\rangle),
\end{eqnarray}
where $c_{i\uparrow}$ is the annihilation operator for $d_{xy}$-orbital electrons on the site $i$ with spin $\sigma$ ($\uparrow$ or $\downarrow$). 
The indices $x$, $y$, $+$ and $-$ stand for the $(a,0)$, $(0,a)$, $(a,a)$ and $(-a,a)$ directions of real-space pairing, respectively. 
In the following we will suppress the constants $a$, $ \hbar $ and $ k_B $ by setting them 1. The next nearest hopping amplitude $t'=0.35t$ reproduces the $\gamma$ band of Sr$_2$RuO$_4$. 
 
We can tune the Lifshitz transition mentioned above directly by changing the chemical potential $ \mu $. 
In this way the Fermi surface switches from a electron- to hole-like shape at a critical value of $ \mu = \mu_c = 4 t' = 1.4 t $ shown in Fig. \ref{lattice} right panel. 
At the same time the topology of the superconducting phase changes as we will show below.

\section{Results}
\begin{figure}[b]
\begin{center}
\includegraphics[height=32mm]{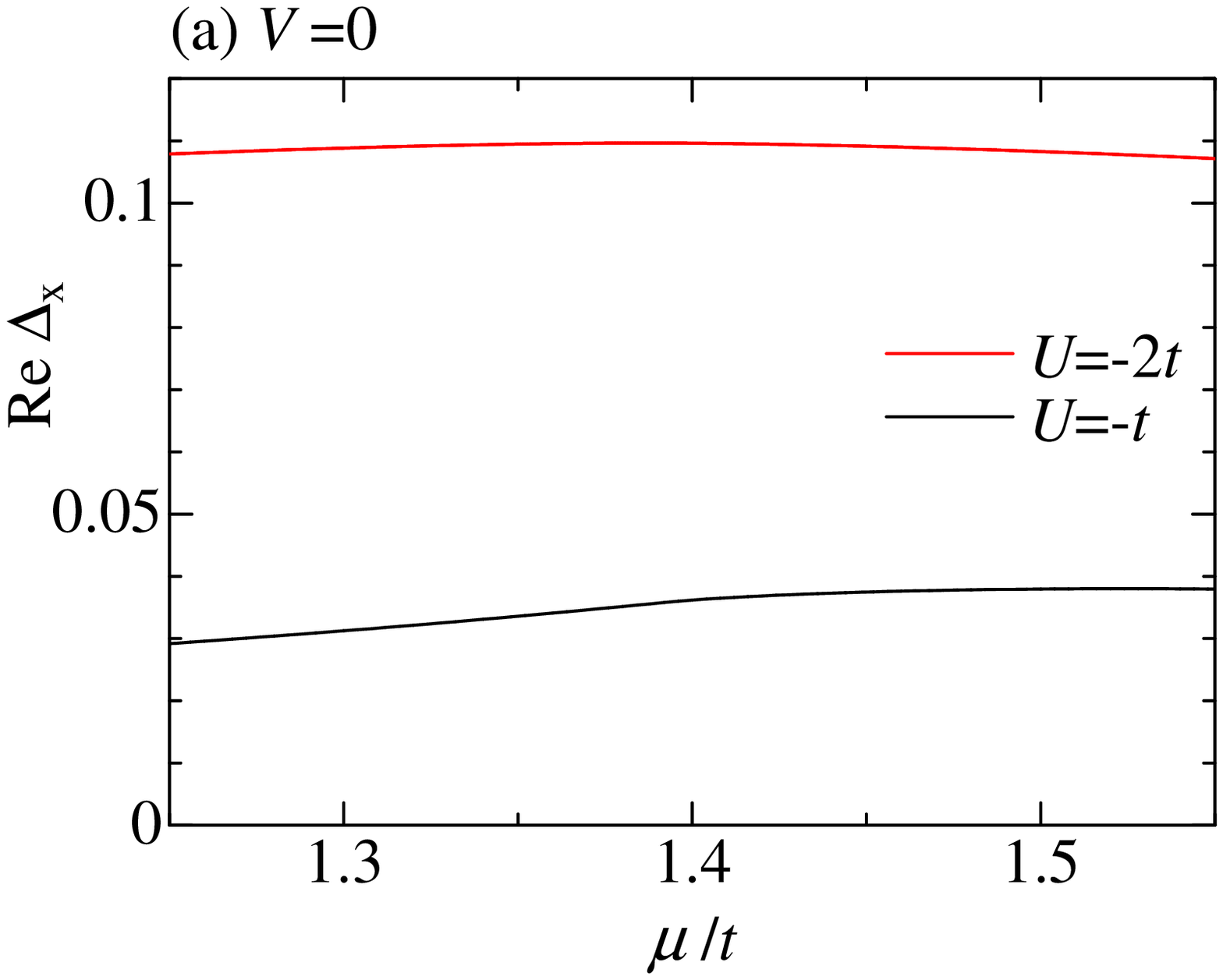}
\includegraphics[height=32mm]{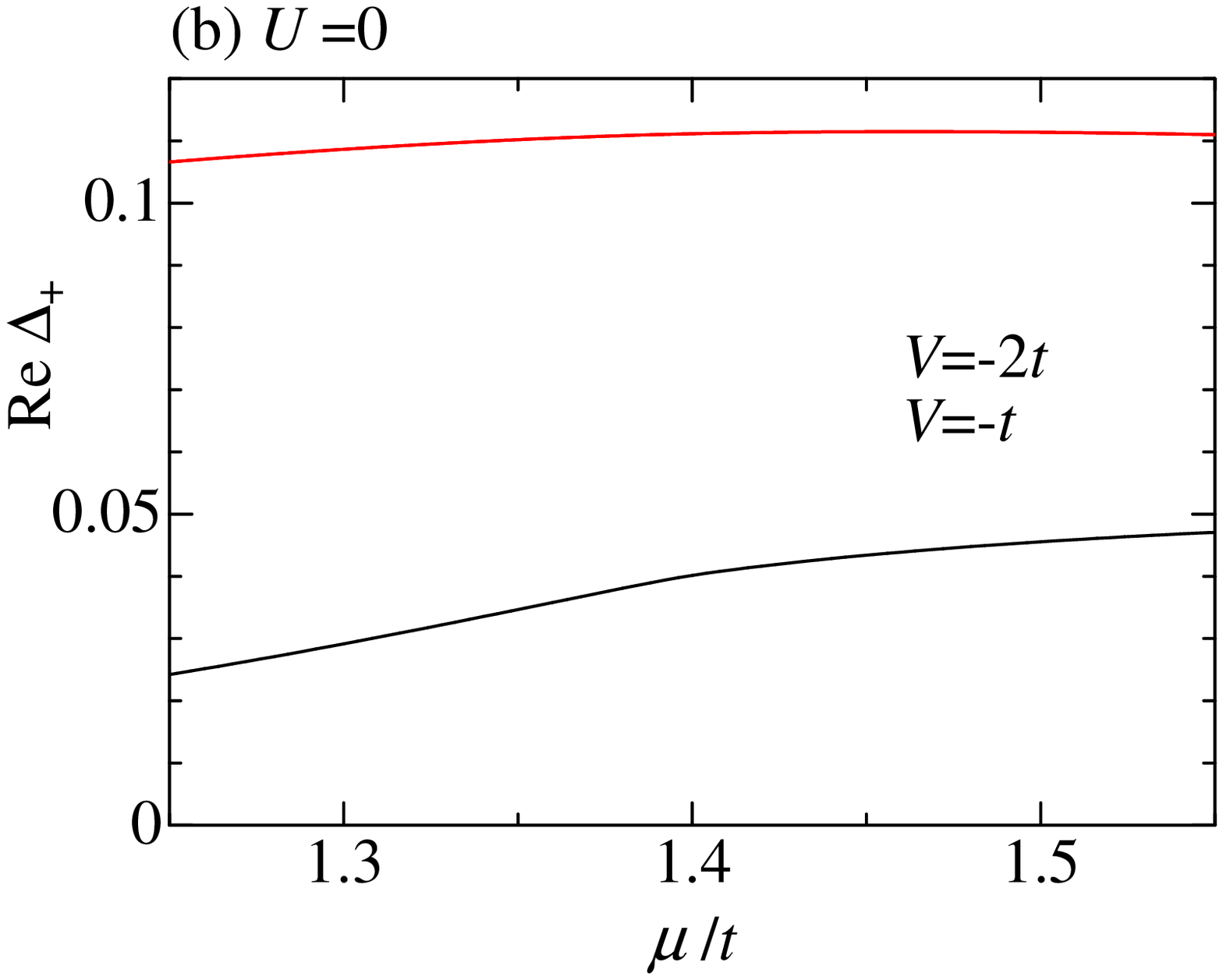}
\caption{(Color online) The order parameters as a function of chemical potential $\mu$ at zero temperature.  The left (right) panel represents Re$\Delta_x$ for $U=-t$, $-2t$ with $V=0$ (Re$\Delta_+$ for $V=-t$, $-2t$ with $U=0$). }
\label{Delta}
\end{center}
\end{figure}

The order parameters $ \Delta_{x,y} $ and $ \Delta_{\pm} $ are determined self-consistently. 
Upon the convergence we obtain the chiral $p$-wave superconducting phase as the most stable state, with $\Delta_y=\pm i\Delta_x$ favored by $U\ne 0$ and $\Delta_-=\pm i\Delta_+$ driven by $V\ne 0$. 
Figure \ref{Delta} shows the order parameters where Im$\Delta_y$ for $U\ne 0$ and $V=0$ (Im$\Delta_-$ for $U= 0$ and $V\ne 0$) is identical to Re$\Delta_x$ (Re$\Delta_+$) with the orbital angular momentum $L_z=+1$. 
Note that the order parameters between the nearest and the next nearest neighbor sites always have same chirality, whose phase difference is $\text{arg}(\Delta_+/\Delta_x)=\pi/4$ when both $U$ and $V$ do not vanish~\cite{imai14}. 
Hereafter we focus on the specific angular momentum $L_z=+1$ state. 
Thus the $\bm d$ vector can be rewritten as $\bm d=\Delta_x\hat{z}(\sin k_x+i \sin k_y)+\Delta_+\hat{z}\{\sin(k_x+k_y)  + i \sin(-k_x+k_y)\}$. 

The bulk quasiparticle spectrum is given by $E_{\bm k\pm}=\pm \sqrt{\varepsilon_{\bm k}^2+|\Delta_{\bm k}|^2}$ and at the van Hove point ($\bm k=(\pi,0)$)  $E_{\bm k\pm}=\pm(4t'-\mu)$, where the gap function $ \Delta_{\bm k} $ vanishes. There are several other discrete points of zero gap within the Brillouin zone. 

\begin{figure}[t]
\begin{center}
\includegraphics[height=40mm]{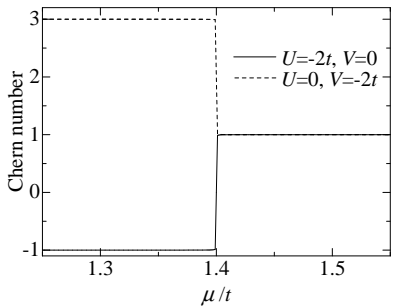}
\caption{The Chern number as a function of chemical potential $\mu$ for $(U, V)=(-2t,0)$ and $(0,-2t)$ at absolute zero temperature. }
\label{NT}
\end{center}
\end{figure}

First we examine the Chern number $N_c$ as a function of the chemical potential $\mu$ for two choices of interactions ($U,V$) at zero temperature, as depicted in Fig. \ref{NT}.  
While for the electron-like Fermi surface the Chern number takes the two values $-1$ and $+3$ for $(U,V)=(-2t,0)$ and $(U,V)=(0,-2t)$ at $\mu<\mu_c$, respectively, it is uniquely $+1$ for $\mu>\mu_c$, the hole-like Fermi surface. This change as function of $ \mu $ defines the Lifshitz transition at $ \mu_c$. Note $\mu_c$ is independent of the magnitude of $U$ and $V$. 
The two values of $ N_c $ for $\mu<\mu_c$ suggest the presence of a topological transition when we continuously interpolate between the two cases of $(U,V)$, yielding a change of $ N_c $ by 4.

We turn now to the thermal transport properties. It has been noticed that for the ordinary thermal current the circulating contribution of the chiral edge states has be considered with care~\cite{qin11}. We will focus here, however, on the thermal Hall conductivity in a chiral $p$-wave superconductor which can be expressed as
\begin{eqnarray}
\kappa_{xy}&=&-\frac{1}{4\pi T}\int d\varepsilon \, \varepsilon^2 \Lambda(\varepsilon) f'(\varepsilon)
\label{eqn:kappa}, \\
\Lambda(\varepsilon)&=&\frac{4\pi}{M}
\sum_{\bm k,n}
{\rm Im}
\left\{
\left \langle \frac{\partial u_{\bm k n}}{\partial k_x} \right| 
\left.\frac{\partial u_{\bm k n}}{\partial k_y} \right\rangle
\right\}
\theta (\varepsilon-E_{\bm k n}), 
\end{eqnarray}
where $f'(\varepsilon)$ is the derivative of the Fermi-distribution function~\cite{sumiyoshi13},  $T$ and $M$ denote temperature and the number of sites, respectively, and $u_{\bm k n}$ ($E_{\bm k n}$) is the periodic part of the Bloch wave function (the eigenvalue) of the BdG equations (Eq. (\ref{eqn:BdG})) for the wave vector $\bm k$ and band index $n$ and is obtained numerically. Note that $\Lambda(0) (\equiv N_{\rm c})$ is identical to the Chern number. 

\begin{figure}[t]
\begin{center}
\includegraphics[height=32mm]{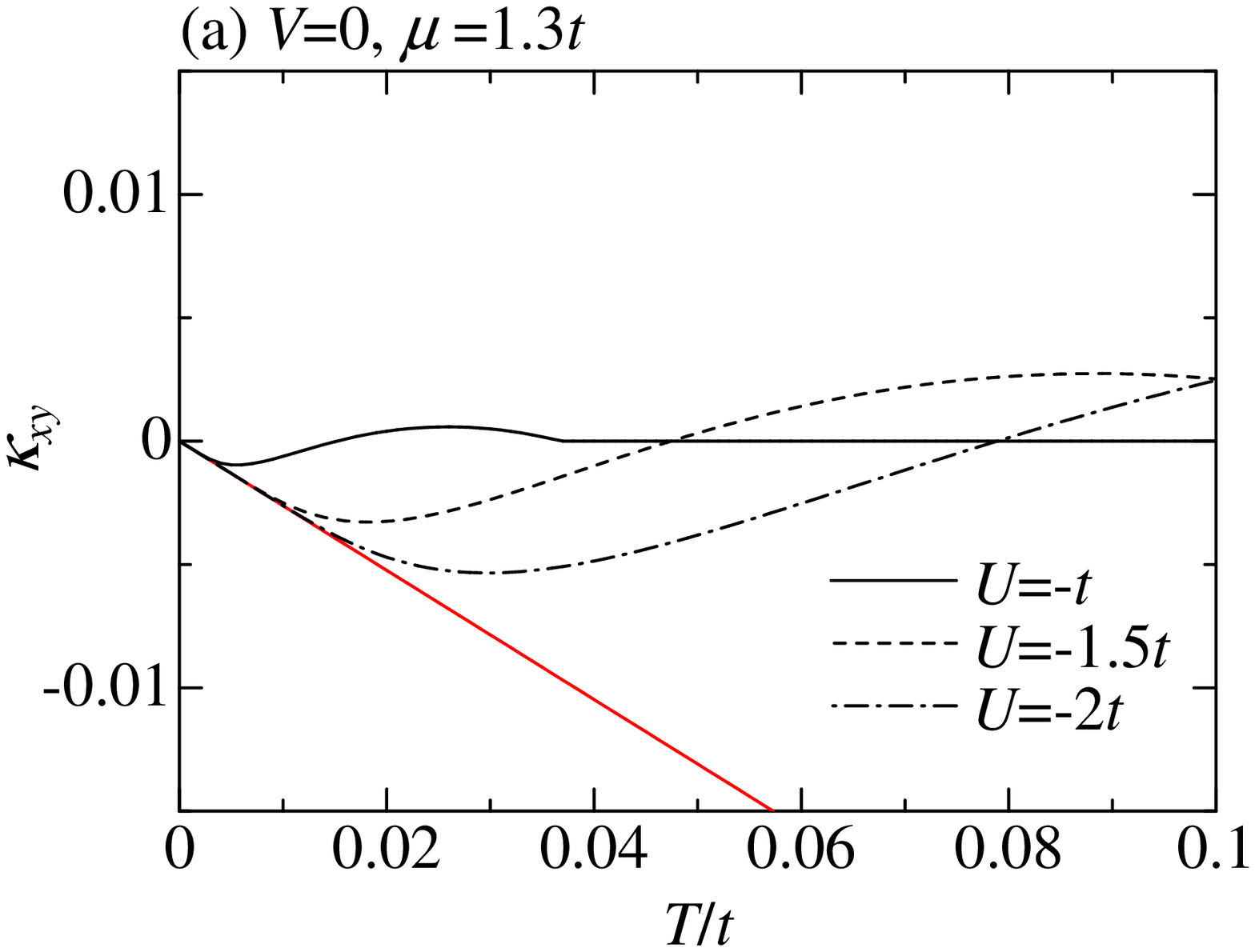}
\includegraphics[height=32mm]{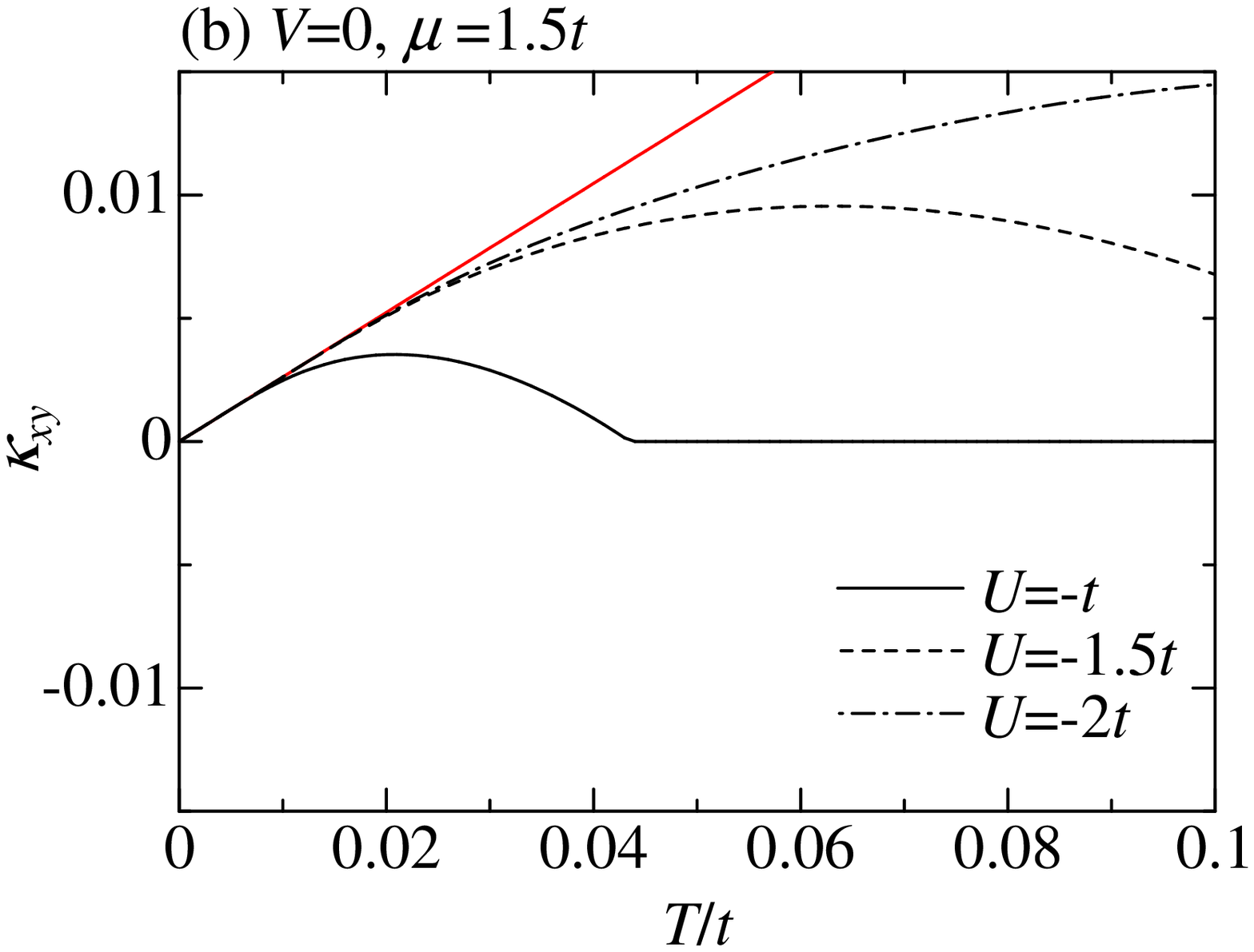}
\includegraphics[height=32mm]{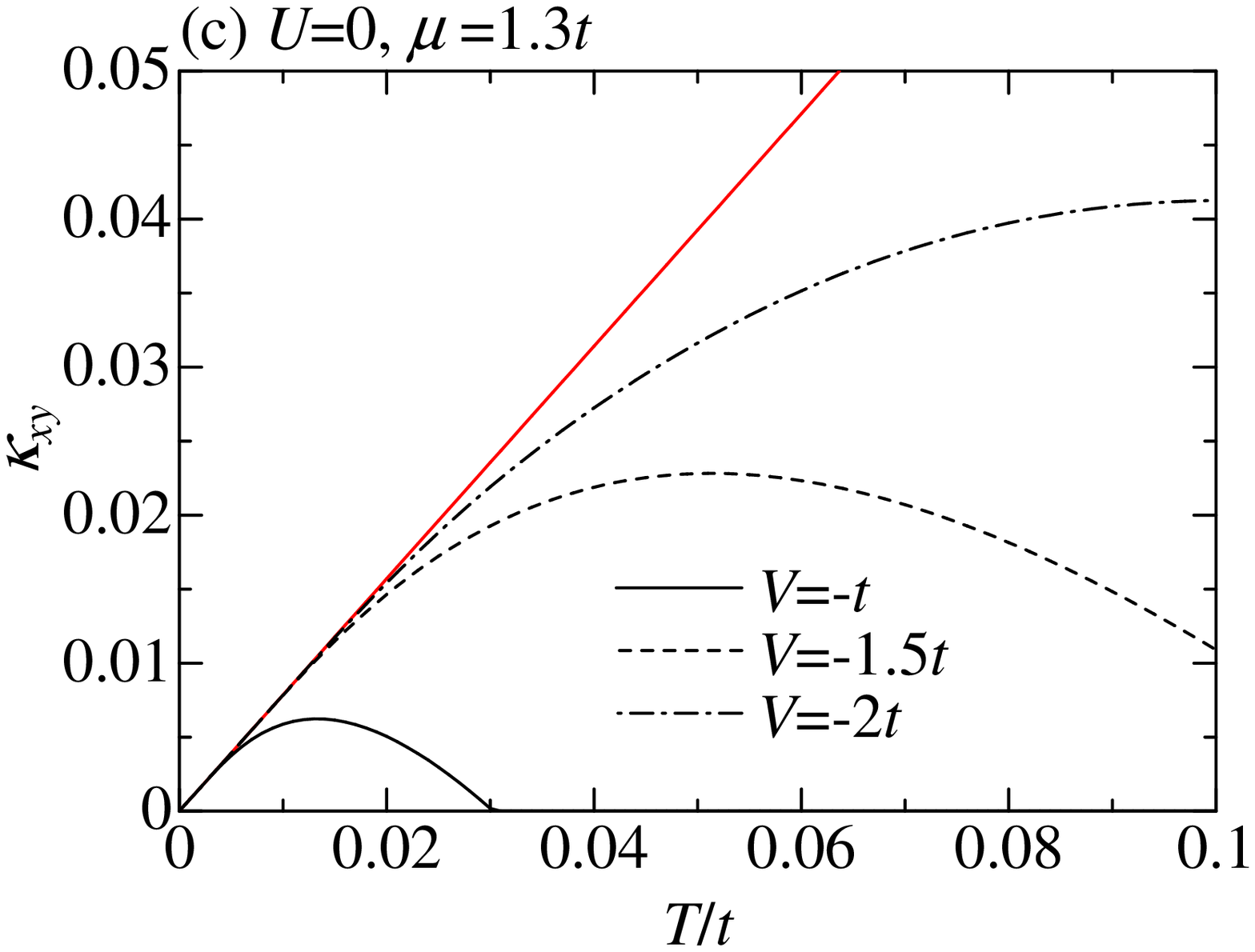}
\includegraphics[height=32mm]{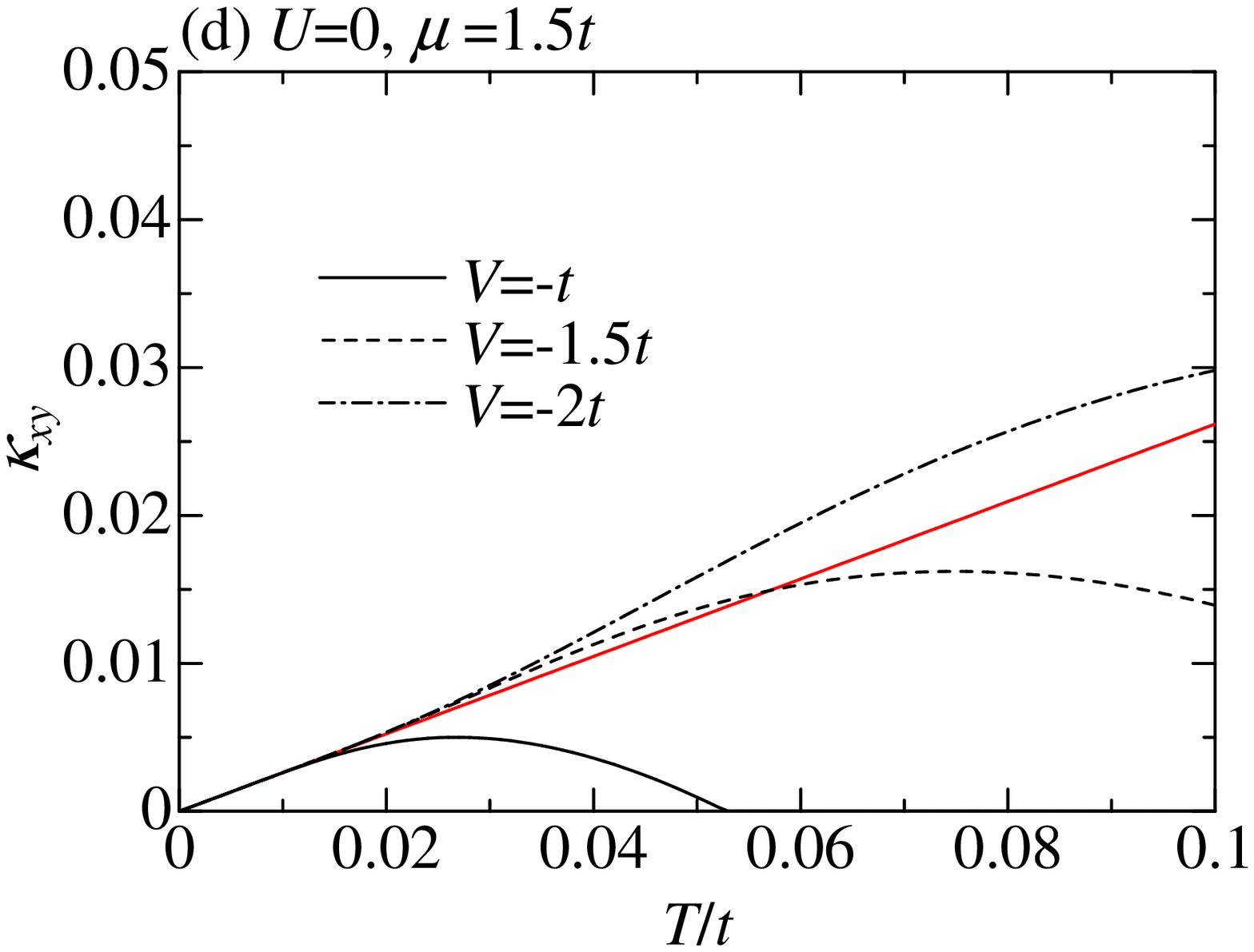}
\caption{(Color online) 
The thermal Hall conductivity $\kappa_{xy}$ as a function of temperature at absolute zero temperature for several choices of $U$ with $V=0$ ((a) and (b)) and several choices of $V$ with $U=0$ ((c) and (d)) at $\mu=1.3t$ (left panels) and $\mu=1.5t$ (right panels). The solid-red line represents $\kappa^{L}_{xy}$ for each set of the parameters ($U$, $V$, $\mu$). 
}
\label{kappa}
\end{center}
\end{figure}

Figure \ref{kappa} displays the temperature dependence of the thermal Hall conductivity. In the low-temperature limit, the thermal Hall conductivity,  $\kappa_{xy}\approx(\pi N_{\rm c}/12) T$~\cite{sumiyoshi13}, is proportional to temperature with a prefactor directly related to the Chern number (the solid-red  lines in Fig.\ref{kappa}). 
This term linear in temperature is defined as 
\begin{eqnarray}
\kappa^{L}_{xy}\equiv \frac{\pi N_{\rm c}}{12}T. 
\end{eqnarray}
The change of the low-$T$ slope of $\kappa_{xy}$ (corresponding to $\kappa^{L}_{xy}$) indicates the change of the Chern number at $\mu=\mu_c$ and conveys that in the very low temperature regime $ \kappa_{xy} $ is uniquely determined by topological properties of the superconducting phase. 

\begin{figure}[b]
\begin{center}
\includegraphics[height=33mm]{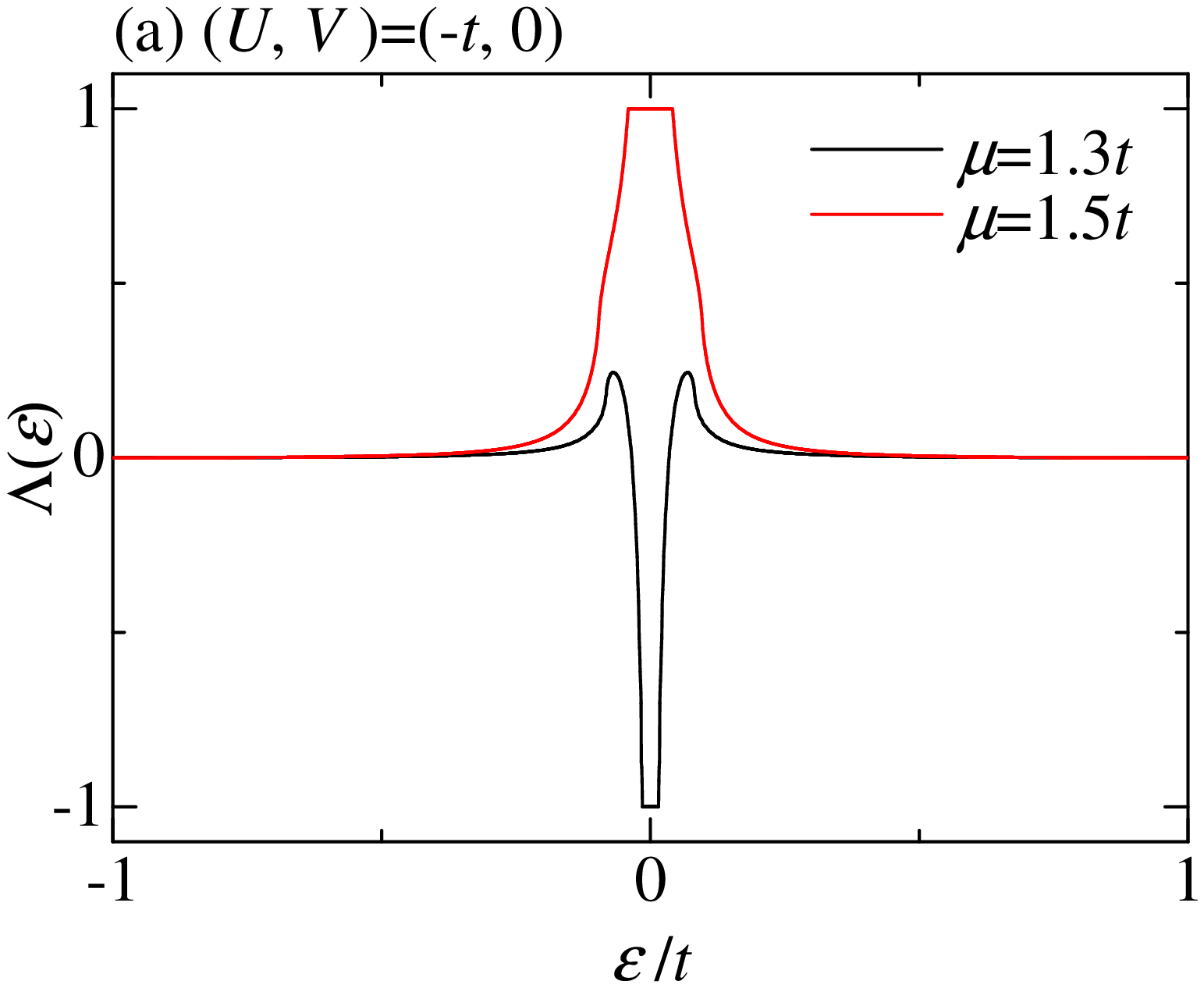}
\includegraphics[height=33mm]{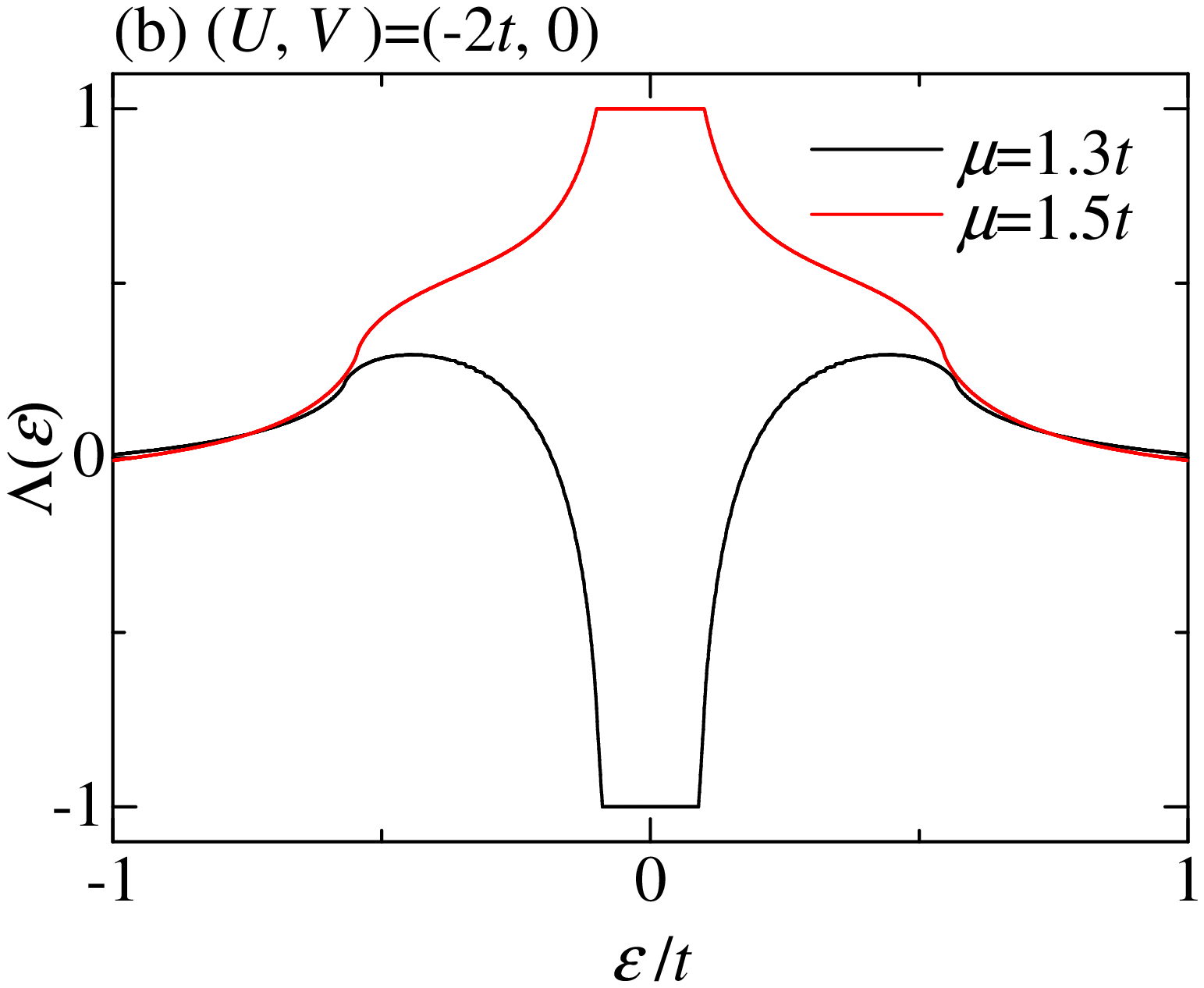}
\includegraphics[height=33mm]{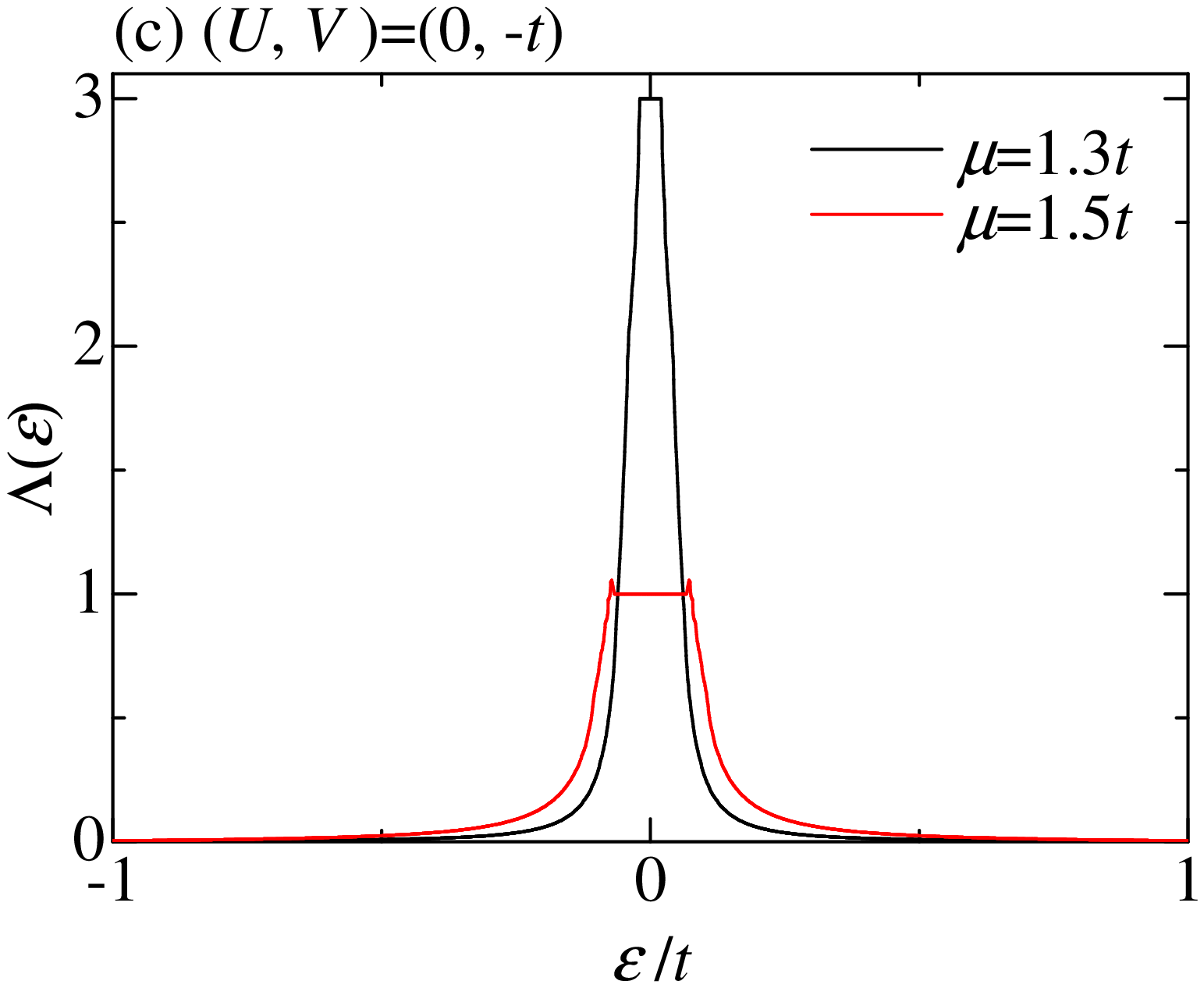}
\includegraphics[height=33mm]{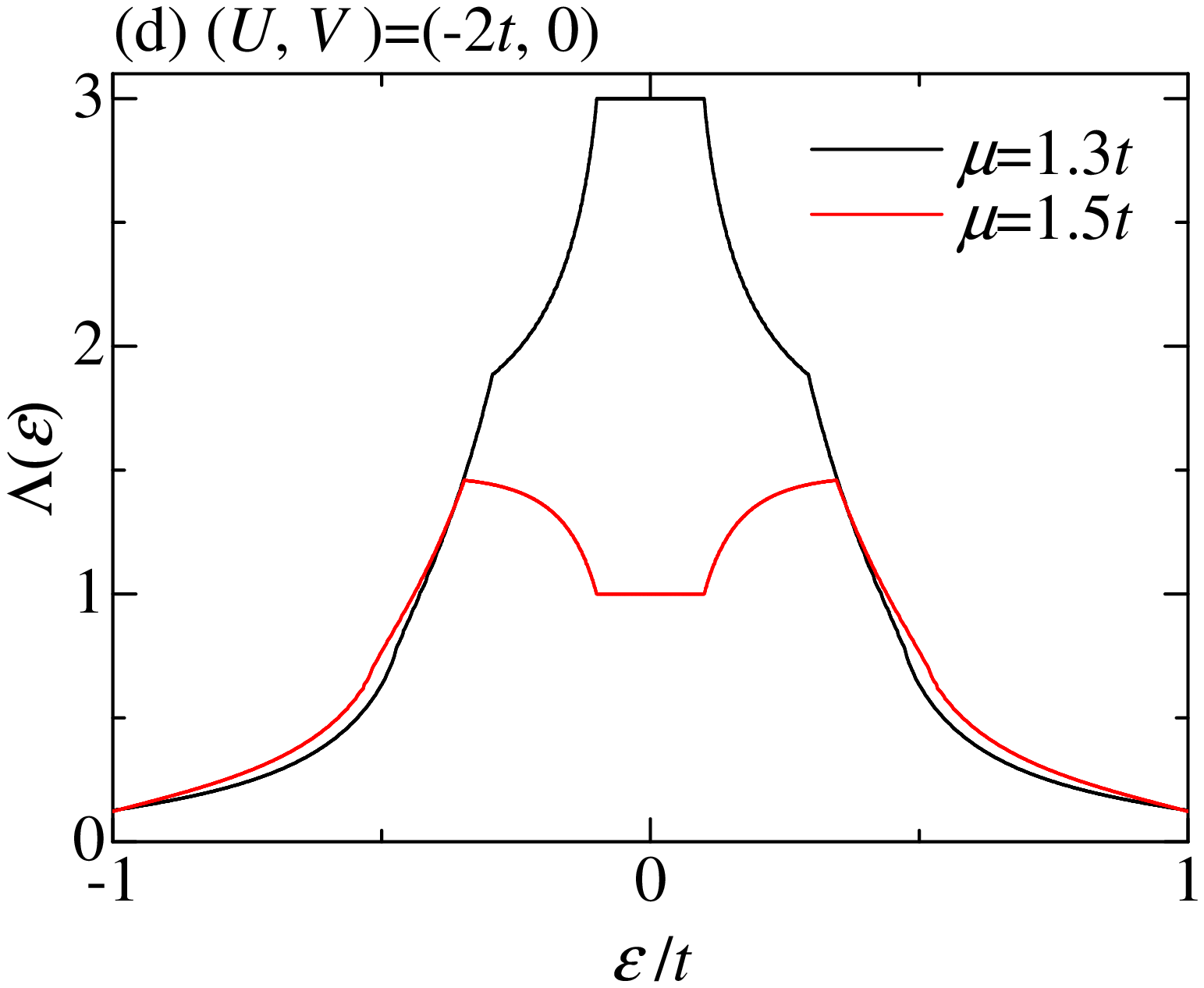}
\caption{(Color online) 
$\Lambda(\varepsilon)$ at zero temperature for $(U,V)=(-t,0),(-2t,0)$ (upper panels) and $(U,V)=(0,-t),(0,-2t)$ (lower panels) at $\mu=1.3t$ (black) and $\mu=1.5t$ (red). 
}
\label{kappa_0}
\end{center}
\end{figure}

We observe a deviation from the $ T$-linear behavior of $ \kappa_{xy} $ in Fig. \ref{kappa} which we trace back to the finite magnitude of the quasiparticle excitation gap, as we will demonstrate in the following. For this purpose we discuss the influence of the structure of $\Lambda(\varepsilon)$ on the temperature dependence of $ \kappa_{xy} $. 
Figure \ref{kappa_0} shows $\Lambda(\varepsilon)$ for several choices of $(U,V)$ at $\mu=1.3t$ and $\mu=1.5t$. 
In our numerical results we find $\Lambda(\varepsilon) = N_c$ in a region $|\varepsilon|\leq \varepsilon_0$ and a rapid change beyond. By comparison we can connect $ \varepsilon_0 $ with the quasiparticle gap $ E_g $ (the lowest excitation energy), i.e. $ \varepsilon_0 = E_g/2 $. For $ | \varepsilon | \gg \varepsilon_0 $ the value of $ \Lambda(\varepsilon) $ shrinks towards zero. This part of the function depends on details of the model. 

For an illustrative approximation of $ \kappa_{xy} $ as a function of temperature we use a piece-wise constant (box shaped) function $\Lambda'(\varepsilon) $ instead of the exact $\Lambda(\varepsilon) $, 
\begin{eqnarray}
\Lambda'(\varepsilon)=\left\{
\begin{array}{cc}
N_c&|\varepsilon | \leq \varepsilon_0 \\
0& {\rm otherwise}
\end{array}
\right..
\end{eqnarray}
Then the evaluation of Eq. (\ref{eqn:kappa}) is straightforward and shows the essential behavior of approximate thermal Hall conductivity $\kappa_{xy}'$ with  $\Lambda'(\varepsilon)$,  
\begin{eqnarray}
\kappa_{xy}'&=&\frac{N_{\rm c}T}{2\pi}
\left\{ 
\frac{\pi^2}{6}
-\gamma(T) e^{-\frac{\varepsilon_0}{T}} 
\right\}
\label{eqn:kappar}
, \\
\gamma(T)&\equiv&
 4\left( \frac{\varepsilon_0}{2T}\right)^2
+4\left( \frac{\varepsilon_0}{2T}\right)+2.  
\label{eqn:kappa_r}
\end{eqnarray}
The second term in Eq. (\ref{eqn:kappar}) is obviously a correction to $ \kappa_{xy}^{L} $ due to contributions of thermally activated quasiparticles and is only a valid approximation as long as $ \varepsilon_0 \gg T $, also in view of
the temperature dependence of the gap which shrinks as temperature increases. 

\begin{figure}[t]
\begin{center}
\includegraphics[height=33mm]{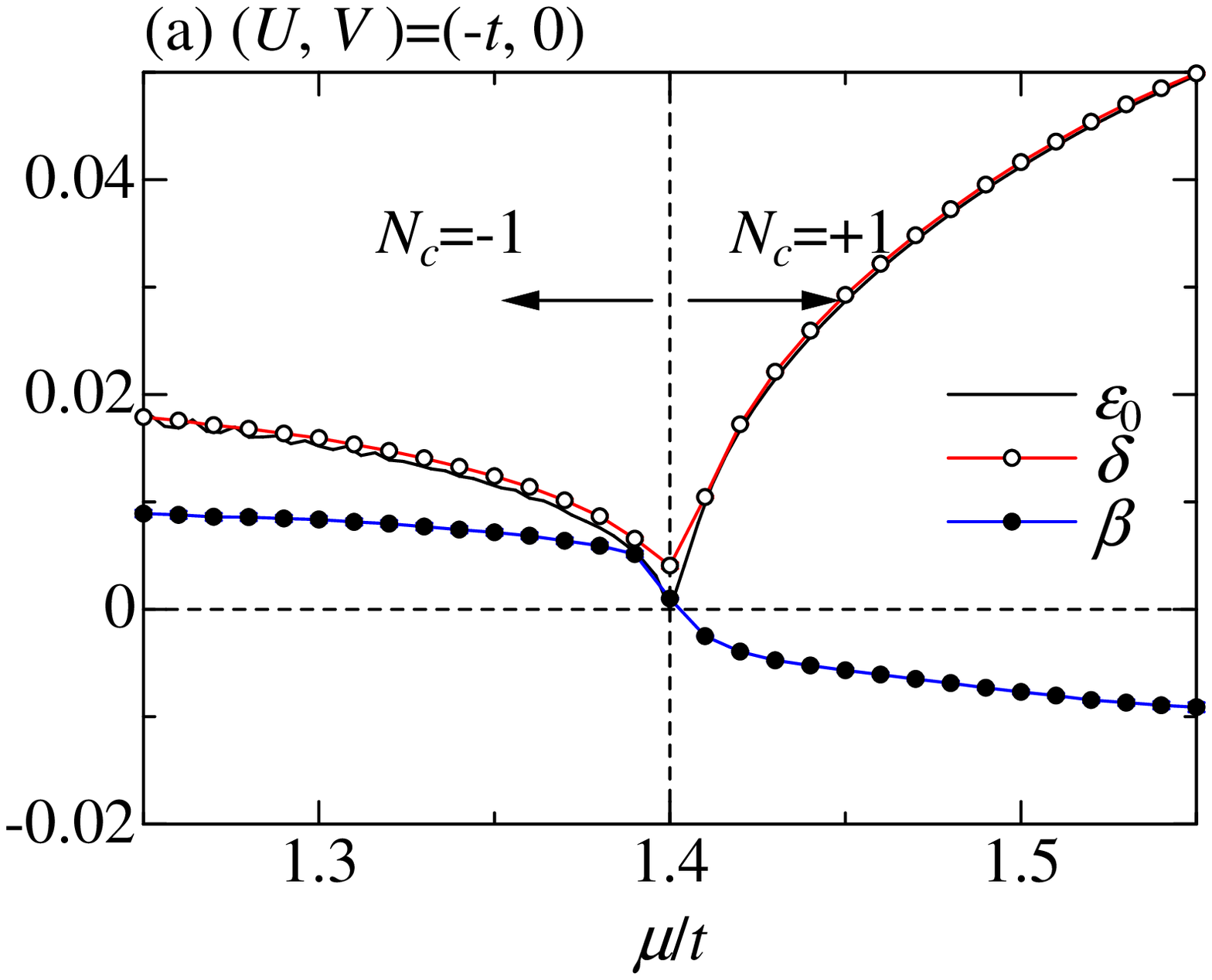}
\includegraphics[height=33mm]{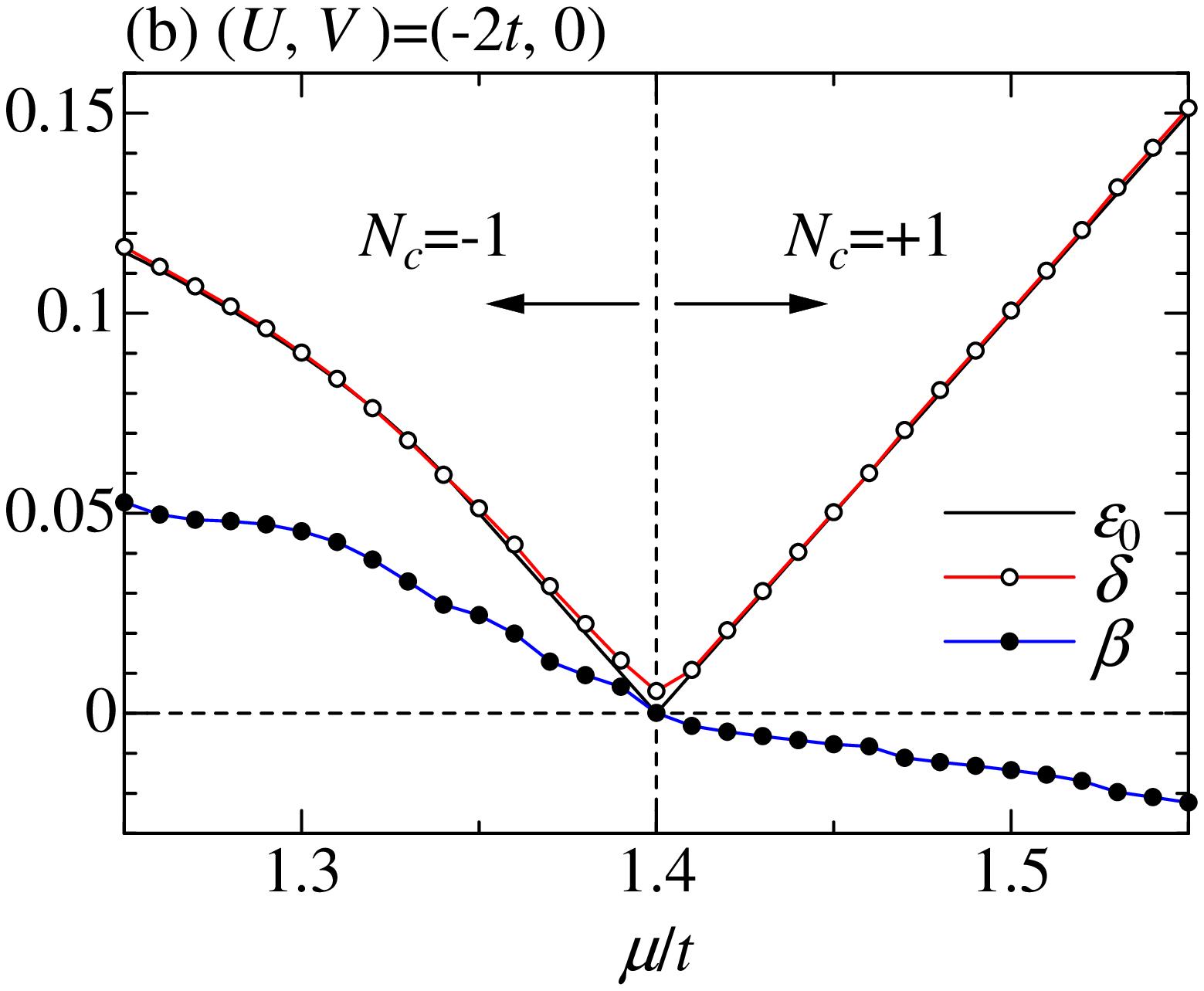}
\includegraphics[height=33mm]{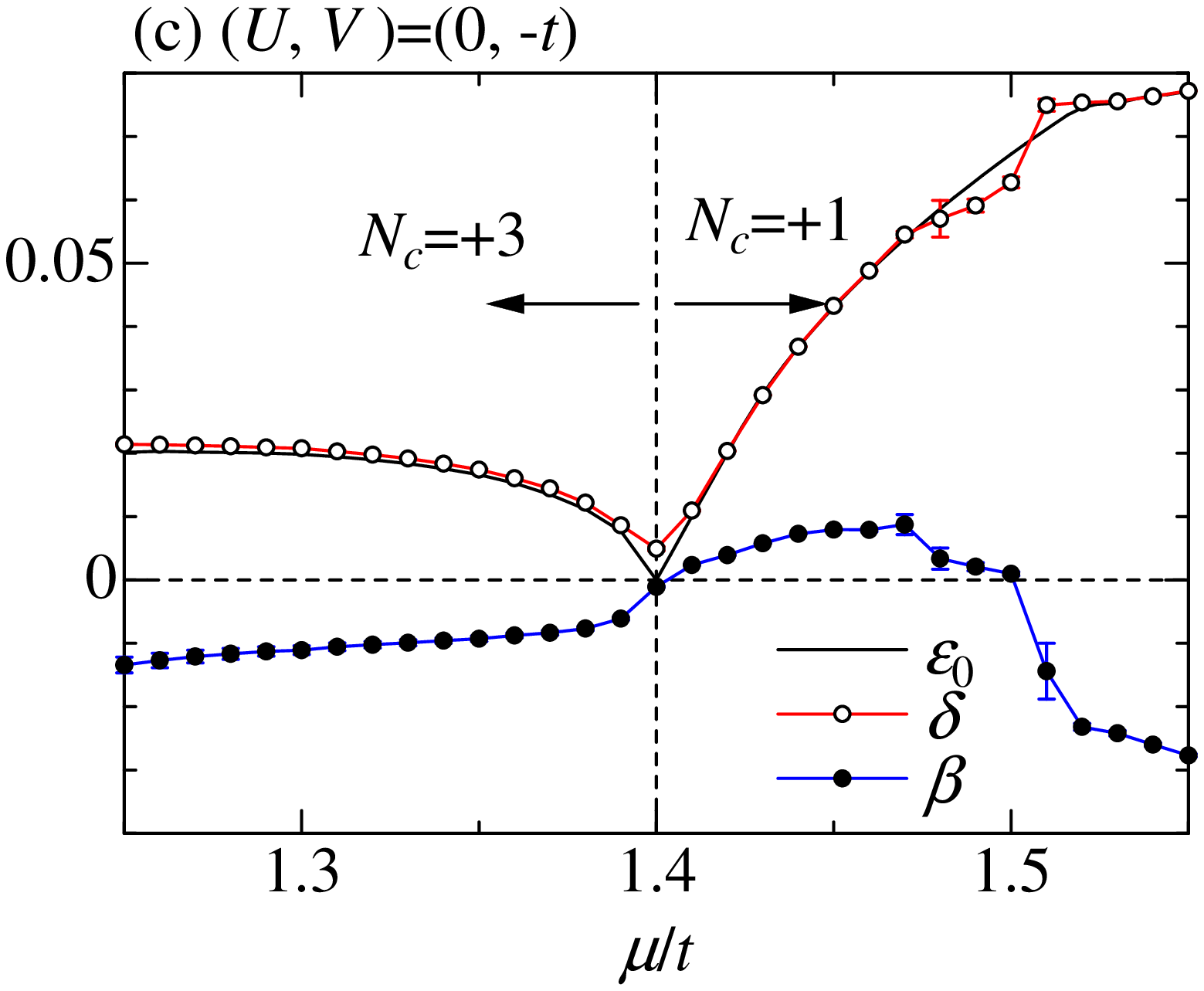}
\includegraphics[height=33mm]{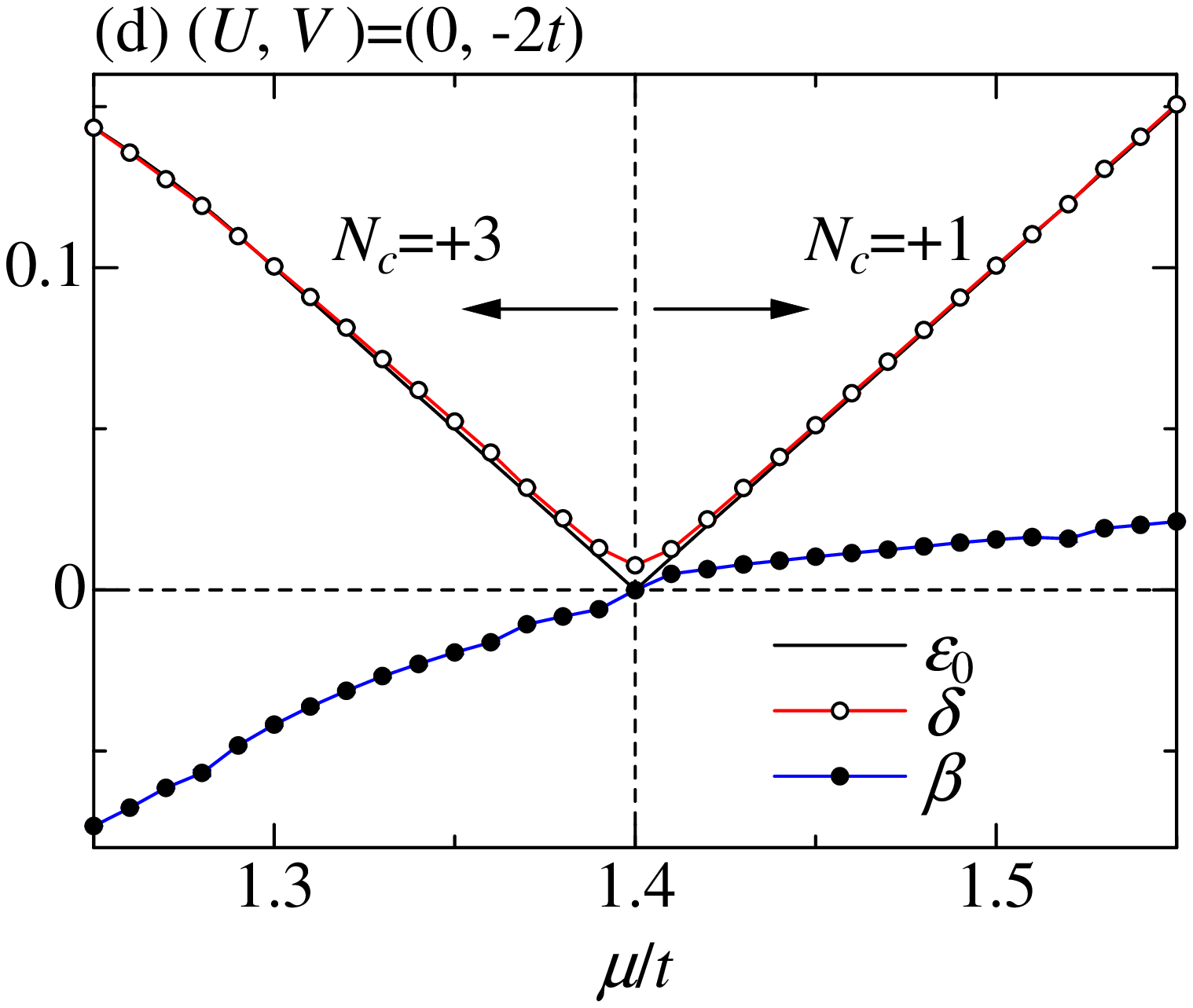}
\caption{(Color online) 
The fitting parameters $\delta$ and $\beta$ as a function of chemical potential $\mu$ for $(U,V)=(-t,0),(-2t,0)$  (upper panels) and $(U,V)=(0,-t),(0,-2t)$ (lower panels). The solid lines stand for the half energy gap $\epsilon_0$ in all panels. 
}
\label{fitting}
\end{center}
\end{figure}

We now compare $ \kappa_{xy}(T) $ with our numerical results. To simplify the expression we use here
\begin{eqnarray}
\kappa_{xy}\approx \tilde{\kappa}_{xy}\equiv\kappa_{xy}^{L}+\beta e^{-\frac{\delta}{T}}, 
\end{eqnarray}
with fitting parameters $\beta$ and $\delta$. Thus, we may use an Arrhenius fit for $\kappa_{xy}-\kappa_{xy}^{L}$. 
Figure \ref{fitting} displays the fitting parameters $\delta$ and $\beta$ for several choices of ($U,V$). 
\begin{figure}[t]
\begin{center}
\includegraphics[height=32mm]{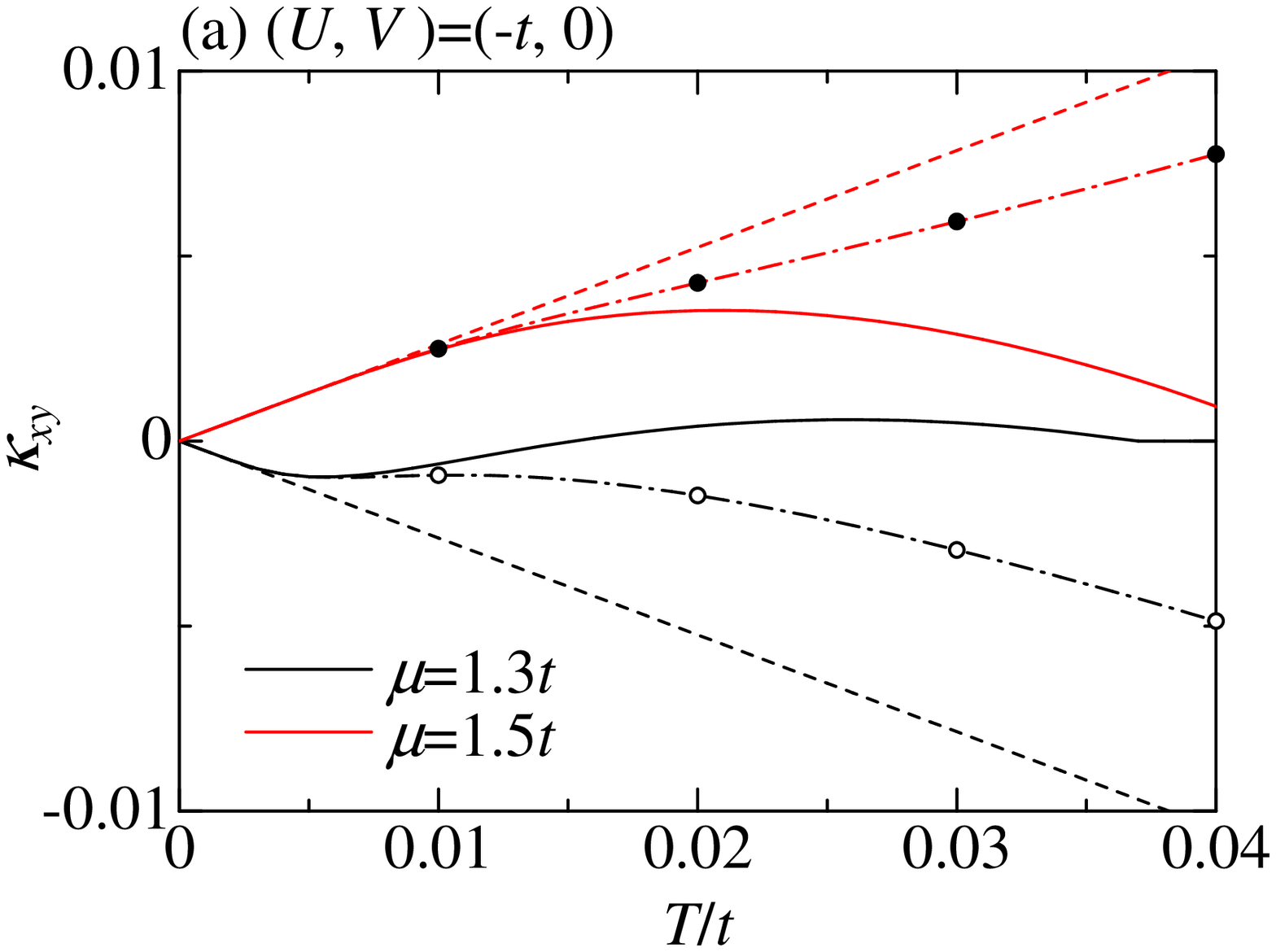}
\includegraphics[height=32mm]{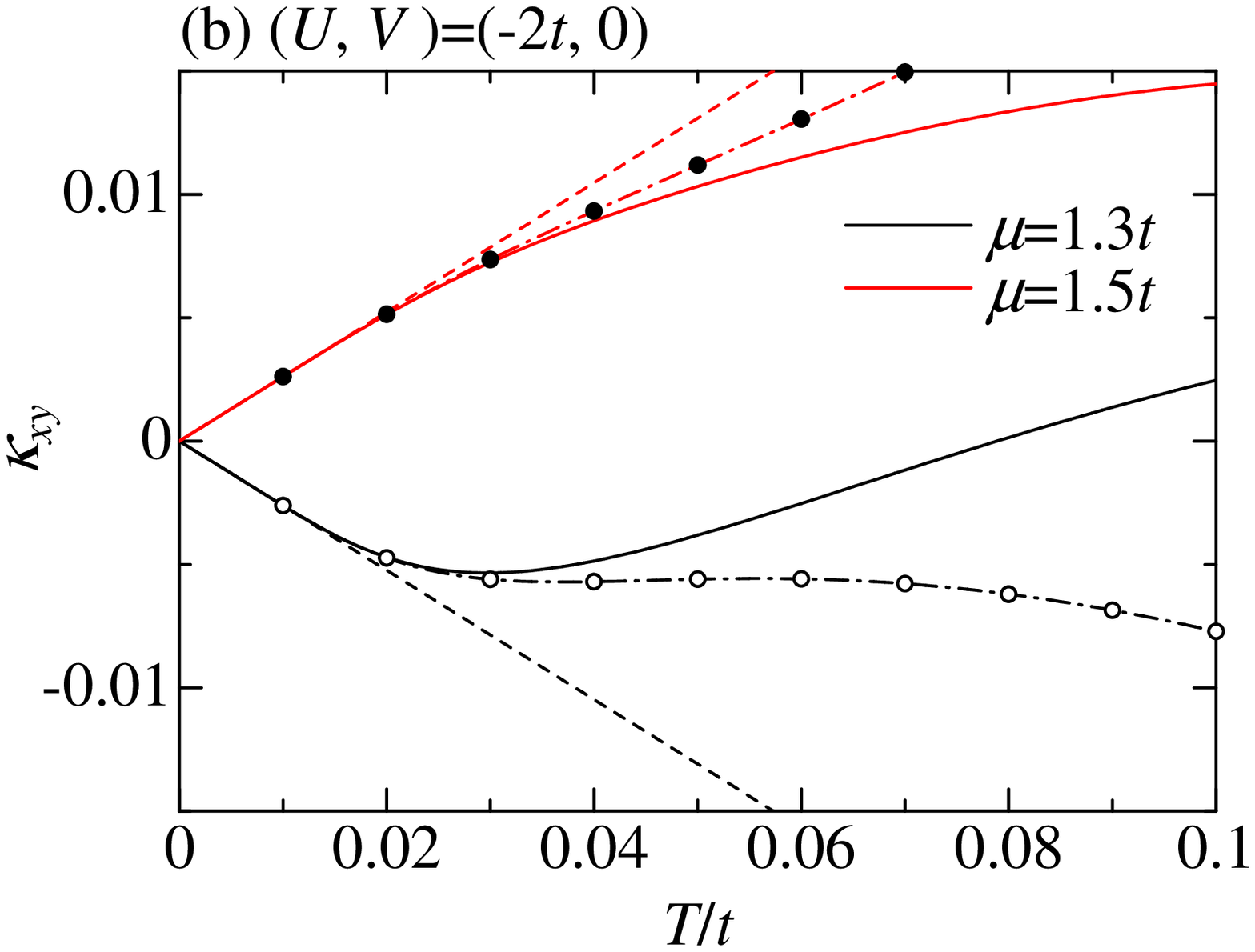}
\includegraphics[height=32mm]{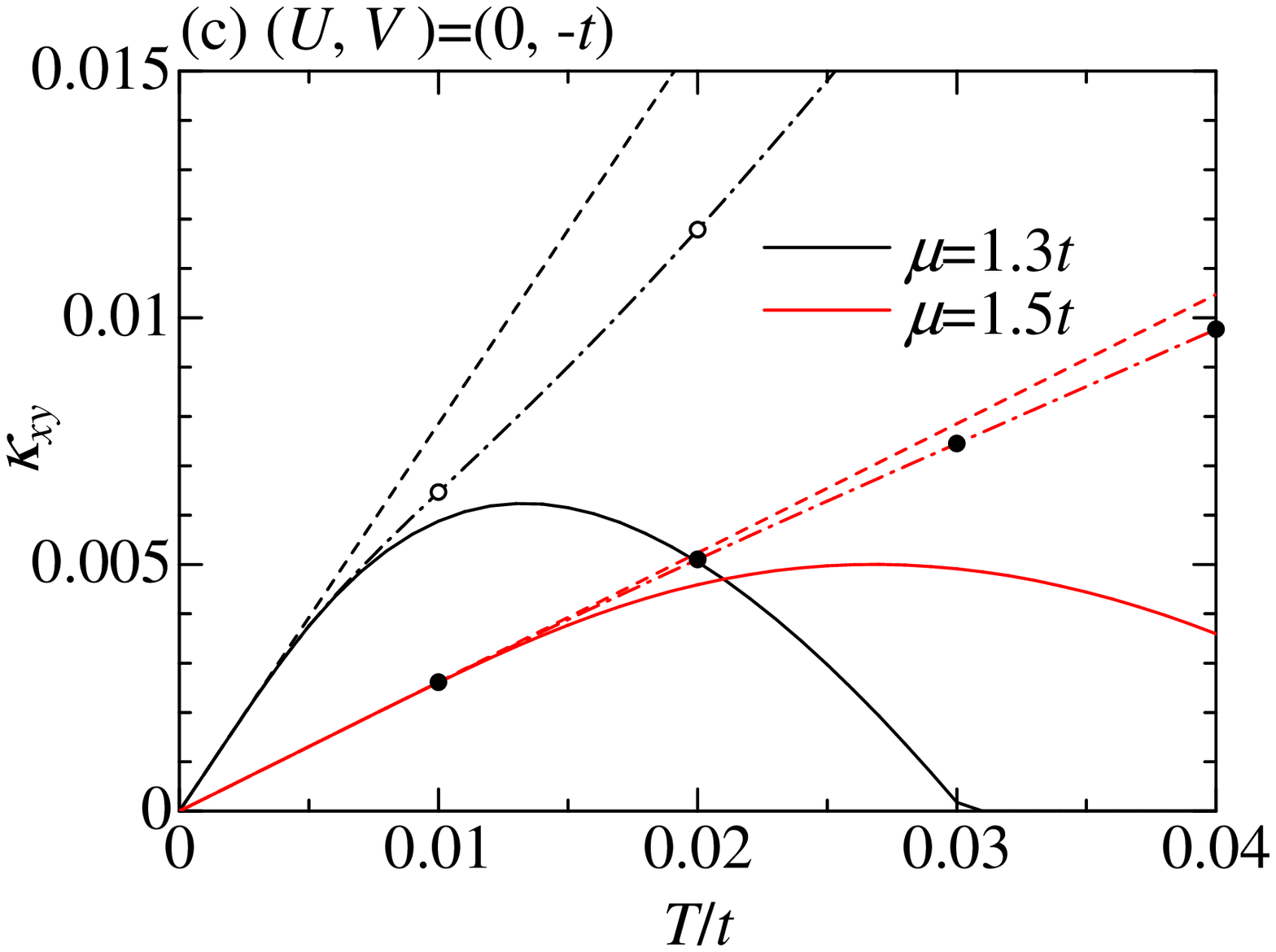}
\includegraphics[height=32mm]{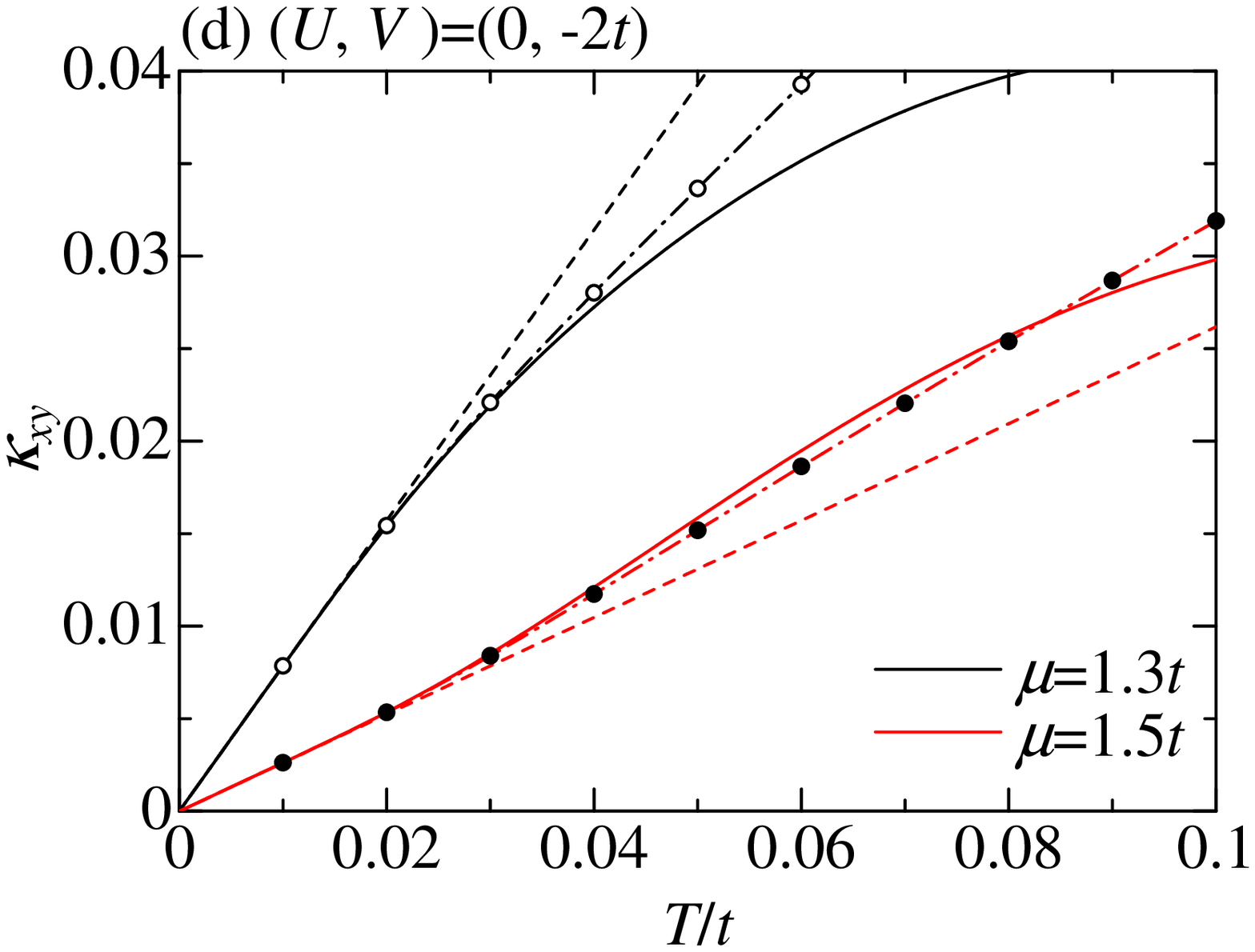}
\caption{(Color online) 
The solid lines, dashed lines and dashed-dotted lines with circles stand for $\kappa_{xy}$, $\kappa_{xy}^{L}$ and $\tilde{\kappa}_{xy}$ as a function of temperature for $(U,V)=(-t,0),(-2t,0)$ (upper panels) and $(U,V)=(0,-t),(0,-2t)$ (lower panels). The black (red) line denotes the results for $\mu=1.3t$ ($\mu=1.5t$). 
}
\label{kappa_fitting}
\end{center}
\end{figure}
In the whole range of $\mu$, $\delta$ is essentially identical with $\varepsilon_0$, where $\varepsilon_0 \approx |4t'-\mu|$ for the Fermi surface close to the van Hove point. Thus, approaching the (topological) Lifshitz transition the quantization becomes thermally softened as the protection due to the quasiparticle gap weakens.

The comparison of $\kappa_{xy}$ with $\tilde{\kappa}_{xy}$ for the estimated $\delta$ and $\beta$ is depicted in Fig. \ref{kappa_fitting} for $\mu=1.3t$ and $\mu=1.5t$. The approximation works well in the temperature range of validity ($ \epsilon_0 \gg T $). 
The parameter $ \beta $ includes information of the shape to $ \Lambda(\varepsilon)$ and determines, in particular,  the sign of the deviation from $T$-linear behavior. 
We find that $\beta $ has the same sign as the Chern number $ N_c $, if $ \Lambda(0) = N_c $ represents a global maximum of $ \Lambda(\varepsilon)$. It is opposite, if $ \Lambda(0) = N_c $ is only a local extremum of $ \Lambda(\varepsilon)$, as it the case, for example, for $ (U,V) = (0,-2t) $ and $ \mu = 1.3 t $ as displayed in Fig. \ref{kappa_0}. Thus, the sign of $ \beta $ depends on whether $ | \Lambda(\varepsilon > \varepsilon_0) | $ is larger or smaller that $ | N_c | $. 
This behavior can be straightforwardly reproduced using Eqs. (\ref{eqn:kappa}) and (\ref{eqn:kappa_r}). 
From the above discussion it becomes clear that  $ \kappa_{xy}(T,\mu)/T $ would display a step-like feature as a function of $ \mu$ at $ \mu=\mu_c $. The deviation from $ \kappa_{xy}^L /T $ would yield a thermal broadening of the step which would be sharp only in the limit of $ T=0 $.

For the purpose of computational feasibility we used rather large interaction strengths, obtaining large gaps as well as short coherence lengths. Our analysis, however, shows that the qualitative behavior remains valid, if these parameters are modified towards more realistic values, because the overall features of $ \Lambda(\varepsilon) $ remain unchanged. An important point is the change of the Chern number appearing also in the plateau of $ \Lambda(\varepsilon) $ around $ \varepsilon =0 $ at the Lifshitz transition, $ \mu = \mu_c $. Here the gap $ \varepsilon_0 $ vanishes as required for a topological transition. 

\section{Conclusions}

Motivated by the spin-triplet superconductor Sr$_2$RuO$_4$, we have investigated the interplay between the thermal Hall effect (Righi-Leduc effect) and the topology of the chiral $p$-wave superconducting phase. We focus on the  $\gamma$ band of Sr$_2$RuO$_4$ which is close to a Lifshitz transition between an electron- and a hole-like Fermi surface,  changing the topological properties of the superconducting phase, in particular, its Chern number. 
The other two Fermi surfaces resulting from the $\alpha$- and $\beta$-bands are not affected much by the transition. 
Moreover, the Chern numbers of these Fermi surfaces compensate each other to zero and do not give rise to a topology related contribution to the thermal Hall coefficient. 

We show that the Lifshitz transition in the $ \gamma $-band not only changes the Fermi surface topology but also yields a change of the Chern number of the chiral superconducting state. This alters the structure of the quasi-particle edge states. 
While the supercurrents of the edge states are a surprisingly little affected by this Lifshitz transition, our calculations show a  strong signature in the thermal Hall effect related to the change of topology of the superconducting phase. 
The temperature-dependence of the thermal Hall conductivity consists of a $T$-linear whose coefficient is uniquely related to the Chern number and terms exponentially depending on temperature. We could show that this latter correction provides the information on the superconducting gap amplitude of the $\gamma$ band as they are induced by thermal occupation of the quasi-particle states in the bulk continuum. The observation of the thermal Hall effect would be a possible way to follow the changes of topology through the Lifshitz transition which would not be possible by detecting edge supercurrents. 
Our discussion gives a qualitative picture what one could observe at a Lifshitz transition which could be possibly induced by doping or uniaxial stress. Clearly a more accurate prediction would be required to include the other two bands which would contribute to the deviations from the universal low-temperature behavior. It is non-trivial to assess these contributions quantitatively, as little is known of the gap structure of these bands.


\begin{acknowledgments}
We are grateful to A. Bouhon, J. Goryo, Y. Maeno, T. Neupert, T. Saso, A. Schnyder and Y. Yanase for many helpful discussions. 
The work is supported by the Swiss National Science Foundation and  by the Ministry of Education, Culture, Sports,
Science and Technology, Japan.
YI is grateful for hospitality by the Pauli Center for Theoretical Studies of ETH Z\"urich. 
\end{acknowledgments}

\appendix

\bibliography{paper.bib}

\providecommand{\noopsort}[1]{}\providecommand{\singleletter}[1]{#1}%
\begin{thebibliography}{29}%
\makeatletter
\providecommand \@ifxundefined [1]{%
 \@ifx{#1\undefined}
}%
\providecommand \@ifnum [1]{%
 \ifnum #1\expandafter \@firstoftwo
 \else \expandafter \@secondoftwo
 \fi
}%
\providecommand \@ifx [1]{%
 \ifx #1\expandafter \@firstoftwo
 \else \expandafter \@secondoftwo
 \fi
}%
\providecommand \natexlab [1]{#1}%
\providecommand \enquote  [1]{``#1''}%
\providecommand \bibnamefont  [1]{#1}%
\providecommand \bibfnamefont [1]{#1}%
\providecommand \citenamefont [1]{#1}%
\providecommand \href@noop [0]{\@secondoftwo}%
\providecommand \href [0]{\begingroup \@sanitize@url \@href}%
\providecommand \@href[1]{\@@startlink{#1}\@@href}%
\providecommand \@@href[1]{\endgroup#1\@@endlink}%
\providecommand \@sanitize@url [0]{\catcode `\\12\catcode `\$12\catcode
  `\&12\catcode `\#12\catcode `\^12\catcode `\_12\catcode `\%12\relax}%
\providecommand \@@startlink[1]{}%
\providecommand \@@endlink[0]{}%
\providecommand \url  [0]{\begingroup\@sanitize@url \@url }%
\providecommand \@url [1]{\endgroup\@href {#1}{\urlprefix }}%
\providecommand \urlprefix  [0]{URL }%
\providecommand \Eprint [0]{\href }%
\providecommand \doibase [0]{http://dx.doi.org/}%
\providecommand \selectlanguage [0]{\@gobble}%
\providecommand \bibinfo  [0]{\@secondoftwo}%
\providecommand \bibfield  [0]{\@secondoftwo}%
\providecommand \translation [1]{[#1]}%
\providecommand \BibitemOpen [0]{}%
\providecommand \bibitemStop [0]{}%
\providecommand \bibitemNoStop [0]{.\EOS\space}%
\providecommand \EOS [0]{\spacefactor3000\relax}%
\providecommand \BibitemShut  [1]{\csname bibitem#1\endcsname}%
\let\auto@bib@innerbib\@empty
\bibitem [{\citenamefont {Maeno}\ \emph {et~al.}(1994)\citenamefont {Maeno},
  \citenamefont {Hashimoto}, \citenamefont {Yoshida}, \citenamefont
  {Nishizaki}, \citenamefont {Fujita}, \citenamefont {Bednorz},\ and\
  \citenamefont {Lichtenberg}}]{maeno94}%
  \BibitemOpen
  \bibfield  {author} {\bibinfo {author} {\bibfnamefont {Y.}~\bibnamefont
  {Maeno}}, \bibinfo {author} {\bibfnamefont {H.}~\bibnamefont {Hashimoto}},
  \bibinfo {author} {\bibfnamefont {K.}~\bibnamefont {Yoshida}}, \bibinfo
  {author} {\bibfnamefont {S.}~\bibnamefont {Nishizaki}}, \bibinfo {author}
  {\bibfnamefont {T.}~\bibnamefont {Fujita}}, \bibinfo {author} {\bibfnamefont
  {J.~G.}\ \bibnamefont {Bednorz}}, \ and\ \bibinfo {author} {\bibfnamefont
  {F.}~\bibnamefont {Lichtenberg}},\ }\href@noop {} {\bibfield  {journal}
  {\bibinfo  {journal} {Nature}\ }\textbf {\bibinfo {volume} {372}},\ \bibinfo
  {pages} {532} (\bibinfo {year} {1994})}\BibitemShut {NoStop}%
\bibitem [{\citenamefont {Mackenzie}\ and\ \citenamefont
  {Maeno}(2003)}]{mackenzie03}%
  \BibitemOpen
  \bibfield  {author} {\bibinfo {author} {\bibfnamefont {A.~P.}\ \bibnamefont
  {Mackenzie}}\ and\ \bibinfo {author} {\bibfnamefont {Y.}~\bibnamefont
  {Maeno}},\ }\href@noop {} {\bibfield  {journal} {\bibinfo  {journal} {Rev.
  Mod. Phys.}\ }\textbf {\bibinfo {volume} {75}},\ \bibinfo {pages} {657}
  (\bibinfo {year} {2003})}\BibitemShut {NoStop}%
\bibitem [{\citenamefont {Maeno}\ \emph {et~al.}(2012)\citenamefont {Maeno},
  \citenamefont {Kittaka}, \citenamefont {Nomura}, \citenamefont {Yonezawa},\
  and\ \citenamefont {Ishida}}]{maeno12}%
  \BibitemOpen
  \bibfield  {author} {\bibinfo {author} {\bibfnamefont {Y.}~\bibnamefont
  {Maeno}}, \bibinfo {author} {\bibfnamefont {S.}~\bibnamefont {Kittaka}},
  \bibinfo {author} {\bibfnamefont {T.}~\bibnamefont {Nomura}}, \bibinfo
  {author} {\bibfnamefont {S.}~\bibnamefont {Yonezawa}}, \ and\ \bibinfo
  {author} {\bibfnamefont {K.}~\bibnamefont {Ishida}},\ }\href@noop {}
  {\bibfield  {journal} {\bibinfo  {journal} {J. Phys. Soc. Jpn.}\ }\textbf
  {\bibinfo {volume} {81}},\ \bibinfo {pages} {011009} (\bibinfo {year}
  {2012})}\BibitemShut {NoStop}%
\bibitem [{\citenamefont {Ishida}\ \emph {et~al.}(1998)\citenamefont {Ishida},
  \citenamefont {Mukuda}, \citenamefont {Kitaoka}, \citenamefont {Asayama},
  \citenamefont {Mao}, \citenamefont {Mori},\ and\ \citenamefont
  {Maeno}}]{ishida98}%
  \BibitemOpen
  \bibfield  {author} {\bibinfo {author} {\bibfnamefont {K.}~\bibnamefont
  {Ishida}}, \bibinfo {author} {\bibfnamefont {H.}~\bibnamefont {Mukuda}},
  \bibinfo {author} {\bibfnamefont {Y.}~\bibnamefont {Kitaoka}}, \bibinfo
  {author} {\bibfnamefont {K.}~\bibnamefont {Asayama}}, \bibinfo {author}
  {\bibfnamefont {Z.~Q.}\ \bibnamefont {Mao}}, \bibinfo {author} {\bibfnamefont
  {Y.}~\bibnamefont {Mori}}, \ and\ \bibinfo {author} {\bibfnamefont
  {Y.}~\bibnamefont {Maeno}},\ }\href@noop {} {\bibfield  {journal} {\bibinfo
  {journal} {Nature (London)}\ }\textbf {\bibinfo {volume} {396}},\ \bibinfo
  {pages} {658} (\bibinfo {year} {1998})}\BibitemShut {NoStop}%
\bibitem [{\citenamefont {Luke}\ \emph {et~al.}(1998)\citenamefont {Luke},
  \citenamefont {Fudamoto}, \citenamefont {Kojima}, \citenamefont {Larkin},
  \citenamefont {Merrin}, \citenamefont {Nachumi}, \citenamefont {Uemura},
  \citenamefont {Maeno}, \citenamefont {Mao}, \citenamefont {Mori},
  \citenamefont {Nakamura},\ and\ \citenamefont {Sigrist}}]{luke98}%
  \BibitemOpen
  \bibfield  {author} {\bibinfo {author} {\bibfnamefont {G.~M.}\ \bibnamefont
  {Luke}}, \bibinfo {author} {\bibfnamefont {Y.}~\bibnamefont {Fudamoto}},
  \bibinfo {author} {\bibfnamefont {K.~M.}\ \bibnamefont {Kojima}}, \bibinfo
  {author} {\bibfnamefont {M.~I.}\ \bibnamefont {Larkin}}, \bibinfo {author}
  {\bibfnamefont {J.}~\bibnamefont {Merrin}}, \bibinfo {author} {\bibfnamefont
  {B.}~\bibnamefont {Nachumi}}, \bibinfo {author} {\bibfnamefont {Y.~J.}\
  \bibnamefont {Uemura}}, \bibinfo {author} {\bibfnamefont {Y.}~\bibnamefont
  {Maeno}}, \bibinfo {author} {\bibfnamefont {Z.~Q.}\ \bibnamefont {Mao}},
  \bibinfo {author} {\bibfnamefont {Y.}~\bibnamefont {Mori}}, \bibinfo {author}
  {\bibfnamefont {H.}~\bibnamefont {Nakamura}}, \ and\ \bibinfo {author}
  {\bibfnamefont {M.}~\bibnamefont {Sigrist}},\ }\href@noop {} {\bibfield
  {journal} {\bibinfo  {journal} {Nature (London)}\ }\textbf {\bibinfo {volume}
  {394}},\ \bibinfo {pages} {558} (\bibinfo {year} {1998})}\BibitemShut
  {NoStop}%
\bibitem [{\citenamefont {Xia}\ \emph {et~al.}(2006)\citenamefont {Xia},
  \citenamefont {Maeno}, \citenamefont {Beyersdorf}, \citenamefont {Fejer},\
  and\ \citenamefont {Kapitulnik}}]{xia06}%
  \BibitemOpen
  \bibfield  {author} {\bibinfo {author} {\bibfnamefont {J.}~\bibnamefont
  {Xia}}, \bibinfo {author} {\bibfnamefont {Y.}~\bibnamefont {Maeno}}, \bibinfo
  {author} {\bibfnamefont {P.~T.}\ \bibnamefont {Beyersdorf}}, \bibinfo
  {author} {\bibfnamefont {M.~M.}\ \bibnamefont {Fejer}}, \ and\ \bibinfo
  {author} {\bibfnamefont {A.}~\bibnamefont {Kapitulnik}},\ }\href@noop {}
  {\bibfield  {journal} {\bibinfo  {journal} {Phys. Rev. Lett.}\ }\textbf
  {\bibinfo {volume} {97}},\ \bibinfo {pages} {167002} (\bibinfo {year}
  {2006})}\BibitemShut {NoStop}%
\bibitem [{\citenamefont {Rice}\ and\ \citenamefont {Sigrist}(1995)}]{rice95}%
  \BibitemOpen
  \bibfield  {author} {\bibinfo {author} {\bibfnamefont {T.~M.}\ \bibnamefont
  {Rice}}\ and\ \bibinfo {author} {\bibfnamefont {M.}~\bibnamefont {Sigrist}},\
  }\href@noop {} {\bibfield  {journal} {\bibinfo  {journal} {J. Phys. Condens.
  Matter}\ }\textbf {\bibinfo {volume} {7}},\ \bibinfo {pages} {643} (\bibinfo
  {year} {1995})}\BibitemShut {NoStop}%
\bibitem [{\citenamefont {Damascelli}\ \emph {et~al.}(2000)\citenamefont
  {Damascelli}, \citenamefont {Lu}, \citenamefont {Shen}, \citenamefont
  {Armitage}, \citenamefont {Ronning}, \citenamefont {Feng}, \citenamefont
  {Kim}, \citenamefont {Shen}, \citenamefont {Kimura}, \citenamefont {Tokura},
  \citenamefont {Mao},\ and\ \citenamefont {Maeno}}]{damascelli00}%
  \BibitemOpen
  \bibfield  {author} {\bibinfo {author} {\bibfnamefont {A.}~\bibnamefont
  {Damascelli}}, \bibinfo {author} {\bibfnamefont {D.~H.}\ \bibnamefont {Lu}},
  \bibinfo {author} {\bibfnamefont {K.~M.}\ \bibnamefont {Shen}}, \bibinfo
  {author} {\bibfnamefont {N.~P.}\ \bibnamefont {Armitage}}, \bibinfo {author}
  {\bibfnamefont {F.}~\bibnamefont {Ronning}}, \bibinfo {author} {\bibfnamefont
  {D.~L.}\ \bibnamefont {Feng}}, \bibinfo {author} {\bibfnamefont
  {C.}~\bibnamefont {Kim}}, \bibinfo {author} {\bibfnamefont {Z.-X.}\
  \bibnamefont {Shen}}, \bibinfo {author} {\bibfnamefont {T.}~\bibnamefont
  {Kimura}}, \bibinfo {author} {\bibfnamefont {Y.}~\bibnamefont {Tokura}},
  \bibinfo {author} {\bibfnamefont {Z.~Q.}\ \bibnamefont {Mao}}, \ and\
  \bibinfo {author} {\bibfnamefont {Y.}~\bibnamefont {Maeno}},\ }\href@noop {}
  {\bibfield  {journal} {\bibinfo  {journal} {Phys. Rev. Lett.}\ }\textbf
  {\bibinfo {volume} {85}},\ \bibinfo {pages} {5194} (\bibinfo {year}
  {2000})}\BibitemShut {NoStop}%
\bibitem [{\citenamefont {Mackenzie}\ \emph {et~al.}(1996)\citenamefont
  {Mackenzie}, \citenamefont {Julian}, \citenamefont {Diver}, \citenamefont
  {McMullan}, \citenamefont {Ray}, \citenamefont {Lonzarich}, \citenamefont
  {Maeno}, \citenamefont {Nishizaki},\ and\ \citenamefont
  {Fujita}}]{mackenzie96}%
  \BibitemOpen
  \bibfield  {author} {\bibinfo {author} {\bibfnamefont {A.~P.}\ \bibnamefont
  {Mackenzie}}, \bibinfo {author} {\bibfnamefont {S.~R.}\ \bibnamefont
  {Julian}}, \bibinfo {author} {\bibfnamefont {A.~J.}\ \bibnamefont {Diver}},
  \bibinfo {author} {\bibfnamefont {G.~J.}\ \bibnamefont {McMullan}}, \bibinfo
  {author} {\bibfnamefont {M.~P.}\ \bibnamefont {Ray}}, \bibinfo {author}
  {\bibfnamefont {G.~G.}\ \bibnamefont {Lonzarich}}, \bibinfo {author}
  {\bibfnamefont {Y.}~\bibnamefont {Maeno}}, \bibinfo {author} {\bibfnamefont
  {S.}~\bibnamefont {Nishizaki}}, \ and\ \bibinfo {author} {\bibfnamefont
  {T.}~\bibnamefont {Fujita}},\ }\href@noop {} {\bibfield  {journal} {\bibinfo
  {journal} {Phys. Rev. Lett.}\ }\textbf {\bibinfo {volume} {76}},\ \bibinfo
  {pages} {3786} (\bibinfo {year} {1996})}\BibitemShut {NoStop}%
\bibitem [{\citenamefont {Bergemann}\ \emph {et~al.}(2000)\citenamefont
  {Bergemann}, \citenamefont {Julian}, \citenamefont {Mackenzie}, \citenamefont
  {NishiZaki},\ and\ \citenamefont {Maeno}}]{bergemann00}%
  \BibitemOpen
  \bibfield  {author} {\bibinfo {author} {\bibfnamefont {C.}~\bibnamefont
  {Bergemann}}, \bibinfo {author} {\bibfnamefont {S.~R.}\ \bibnamefont
  {Julian}}, \bibinfo {author} {\bibfnamefont {A.~P.}\ \bibnamefont
  {Mackenzie}}, \bibinfo {author} {\bibfnamefont {S.}~\bibnamefont
  {NishiZaki}}, \ and\ \bibinfo {author} {\bibfnamefont {Y.}~\bibnamefont
  {Maeno}},\ }\href@noop {} {\bibfield  {journal} {\bibinfo  {journal} {Phys.
  Rev. Lett.}\ }\textbf {\bibinfo {volume} {84}},\ \bibinfo {pages} {2662}
  (\bibinfo {year} {2000})}\BibitemShut {NoStop}%
\bibitem [{\citenamefont {Wang}\ \emph {et~al.}(2013)\citenamefont {Wang},
  \citenamefont {Platt}, \citenamefont {Yang}, \citenamefont {Honerkamp},
  \citenamefont {Zhang}, \citenamefont {Hanke}, \citenamefont {Rice},\ and\
  \citenamefont {Thomale}}]{wang13}%
  \BibitemOpen
  \bibfield  {author} {\bibinfo {author} {\bibfnamefont {Q.-H.}\ \bibnamefont
  {Wang}}, \bibinfo {author} {\bibfnamefont {C.}~\bibnamefont {Platt}},
  \bibinfo {author} {\bibfnamefont {Y.}~\bibnamefont {Yang}}, \bibinfo {author}
  {\bibfnamefont {C.}~\bibnamefont {Honerkamp}}, \bibinfo {author}
  {\bibfnamefont {F.~C.}\ \bibnamefont {Zhang}}, \bibinfo {author}
  {\bibfnamefont {W.}~\bibnamefont {Hanke}}, \bibinfo {author} {\bibfnamefont
  {T.~M.}\ \bibnamefont {Rice}}, \ and\ \bibinfo {author} {\bibfnamefont
  {R.}~\bibnamefont {Thomale}},\ }\href@noop {} {\bibfield  {journal} {\bibinfo
   {journal} {Eur. Phys. Lett.}\ }\textbf {\bibinfo {volume} {104}},\ \bibinfo
  {pages} {17013} (\bibinfo {year} {2013})}\BibitemShut {NoStop}%
\bibitem [{\citenamefont {Imai}\ \emph {et~al.}(2013)\citenamefont {Imai},
  \citenamefont {Wakabayashi},\ and\ \citenamefont {Sigrist}}]{imai13}%
  \BibitemOpen
  \bibfield  {author} {\bibinfo {author} {\bibfnamefont {Y.}~\bibnamefont
  {Imai}}, \bibinfo {author} {\bibfnamefont {K.}~\bibnamefont {Wakabayashi}}, \
  and\ \bibinfo {author} {\bibfnamefont {M.}~\bibnamefont {Sigrist}},\
  }\href@noop {} {\bibfield  {journal} {\bibinfo  {journal} {Phys. Rev. B}\
  }\textbf {\bibinfo {volume} {88}},\ \bibinfo {pages} {144503} (\bibinfo
  {year} {2013})}\BibitemShut {NoStop}%
\bibitem [{\citenamefont {Matzdorf}\ \emph {et~al.}(2000)\citenamefont
  {Matzdorf}, \citenamefont {Fang}, \citenamefont {Ismail}, \citenamefont
  {Zhang}, \citenamefont {Kimura}, \citenamefont {Tokura}, \citenamefont
  {Terakura},\ and\ \citenamefont {Plummer}}]{Matzdorf00}%
  \BibitemOpen
  \bibfield  {author} {\bibinfo {author} {\bibfnamefont {R.}~\bibnamefont
  {Matzdorf}}, \bibinfo {author} {\bibfnamefont {Z.}~\bibnamefont {Fang}},
  \bibinfo {author} {\bibnamefont {Ismail}}, \bibinfo {author} {\bibfnamefont
  {J.~D.}\ \bibnamefont {Zhang}}, \bibinfo {author} {\bibfnamefont
  {T.}~\bibnamefont {Kimura}}, \bibinfo {author} {\bibfnamefont
  {Y.}~\bibnamefont {Tokura}}, \bibinfo {author} {\bibfnamefont
  {K.}~\bibnamefont {Terakura}}, \ and\ \bibinfo {author} {\bibfnamefont
  {E.~W.}\ \bibnamefont {Plummer}},\ }\href@noop {} {\bibfield  {journal}
  {\bibinfo  {journal} {Science}\ }\textbf {\bibinfo {volume} {289}},\ \bibinfo
  {pages} {746} (\bibinfo {year} {2000})}\BibitemShut {NoStop}%
\bibitem [{\citenamefont {Veenstra}\ \emph {et~al.}(2013)\citenamefont
  {Veenstra}, \citenamefont {Zhu}, \citenamefont {Ludbrook}, \citenamefont
  {Capsoni}, \citenamefont {Levy}, \citenamefont {Nicolaou}, \citenamefont
  {Rosen}, \citenamefont {Comin}, \citenamefont {Kittaka}, \citenamefont
  {Maeno}, \citenamefont {Elfimov},\ and\ \citenamefont
  {Damascelli}}]{veenstra13}%
  \BibitemOpen
  \bibfield  {author} {\bibinfo {author} {\bibfnamefont {C.~N.}\ \bibnamefont
  {Veenstra}}, \bibinfo {author} {\bibfnamefont {Z.-H.}\ \bibnamefont {Zhu}},
  \bibinfo {author} {\bibfnamefont {B.}~\bibnamefont {Ludbrook}}, \bibinfo
  {author} {\bibfnamefont {M.}~\bibnamefont {Capsoni}}, \bibinfo {author}
  {\bibfnamefont {G.}~\bibnamefont {Levy}}, \bibinfo {author} {\bibfnamefont
  {A.}~\bibnamefont {Nicolaou}}, \bibinfo {author} {\bibfnamefont {J.~A.}\
  \bibnamefont {Rosen}}, \bibinfo {author} {\bibfnamefont {R.}~\bibnamefont
  {Comin}}, \bibinfo {author} {\bibfnamefont {S.}~\bibnamefont {Kittaka}},
  \bibinfo {author} {\bibfnamefont {Y.}~\bibnamefont {Maeno}}, \bibinfo
  {author} {\bibfnamefont {I.~S.}\ \bibnamefont {Elfimov}}, \ and\ \bibinfo
  {author} {\bibfnamefont {A.}~\bibnamefont {Damascelli}},\ }\href@noop {}
  {\bibfield  {journal} {\bibinfo  {journal} {Phys. Rev. Lett.}\ }\textbf
  {\bibinfo {volume} {110}},\ \bibinfo {pages} {097004} (\bibinfo {year}
  {2013})}\BibitemShut {NoStop}%
\bibitem [{\citenamefont {Imai}\ \emph {et~al.}(2014)\citenamefont {Imai},
  \citenamefont {Wakabayashi},\ and\ \citenamefont {Sigrist}}]{imai14}%
  \BibitemOpen
  \bibfield  {author} {\bibinfo {author} {\bibfnamefont {Y.}~\bibnamefont
  {Imai}}, \bibinfo {author} {\bibfnamefont {K.}~\bibnamefont {Wakabayashi}}, \
  and\ \bibinfo {author} {\bibfnamefont {M.}~\bibnamefont {Sigrist}},\
  }\href@noop {} {\bibfield  {journal} {\bibinfo  {journal} {J. Phys. Soc.
  Jpn}\ }\textbf {\bibinfo {volume} {83}},\ \bibinfo {pages} {124712} (\bibinfo
  {year} {2014})}\BibitemShut {NoStop}%
\bibitem [{\citenamefont {Kikugawa}\ \emph
  {et~al.}(2004{\natexlab{a}})\citenamefont {Kikugawa}, \citenamefont
  {Mackenzie}, \citenamefont {Bergemann}, \citenamefont {Borzi}, \citenamefont
  {Grigera},\ and\ \citenamefont {Maeno}}]{kikugawa04A}%
  \BibitemOpen
  \bibfield  {author} {\bibinfo {author} {\bibfnamefont {N.}~\bibnamefont
  {Kikugawa}}, \bibinfo {author} {\bibfnamefont {A.~P.}\ \bibnamefont
  {Mackenzie}}, \bibinfo {author} {\bibfnamefont {C.}~\bibnamefont
  {Bergemann}}, \bibinfo {author} {\bibfnamefont {R.~A.}\ \bibnamefont
  {Borzi}}, \bibinfo {author} {\bibfnamefont {S.~A.}\ \bibnamefont {Grigera}},
  \ and\ \bibinfo {author} {\bibfnamefont {Y.}~\bibnamefont {Maeno}},\
  }\href@noop {} {\bibfield  {journal} {\bibinfo  {journal} {Phys. Rev. B}\
  }\textbf {\bibinfo {volume} {70}},\ \bibinfo {pages} {060508(R)} (\bibinfo
  {year} {2004}{\natexlab{a}})}\BibitemShut {NoStop}%
\bibitem [{\citenamefont {Kikugawa}\ \emph
  {et~al.}(2004{\natexlab{b}})\citenamefont {Kikugawa}, \citenamefont
  {Bergemann}, \citenamefont {Mackenzie},\ and\ \citenamefont
  {Maeno}}]{kikugawa04B}%
  \BibitemOpen
  \bibfield  {author} {\bibinfo {author} {\bibfnamefont {N.}~\bibnamefont
  {Kikugawa}}, \bibinfo {author} {\bibfnamefont {C.}~\bibnamefont {Bergemann}},
  \bibinfo {author} {\bibfnamefont {A.~P.}\ \bibnamefont {Mackenzie}}, \ and\
  \bibinfo {author} {\bibfnamefont {Y.}~\bibnamefont {Maeno}},\ }\href@noop {}
  {\bibfield  {journal} {\bibinfo  {journal} {Phys. Rev. B}\ }\textbf {\bibinfo
  {volume} {70}},\ \bibinfo {pages} {134520} (\bibinfo {year}
  {2004}{\natexlab{b}})}\BibitemShut {NoStop}%
\bibitem [{\citenamefont {Shen}\ \emph {et~al.}(2007)\citenamefont {Shen},
  \citenamefont {Kikugawa}, \citenamefont {Bergemann}, \citenamefont {Balicas},
  \citenamefont {Baumberger}, \citenamefont {Meevasana}, \citenamefont {Ingle},
  \citenamefont {Maeno}, \citenamefont {Shen},\ and\ \citenamefont
  {Mackenzie}}]{shen07}%
  \BibitemOpen
  \bibfield  {author} {\bibinfo {author} {\bibfnamefont {K.~M.}\ \bibnamefont
  {Shen}}, \bibinfo {author} {\bibfnamefont {N.}~\bibnamefont {Kikugawa}},
  \bibinfo {author} {\bibfnamefont {C.}~\bibnamefont {Bergemann}}, \bibinfo
  {author} {\bibfnamefont {L.}~\bibnamefont {Balicas}}, \bibinfo {author}
  {\bibfnamefont {F.}~\bibnamefont {Baumberger}}, \bibinfo {author}
  {\bibfnamefont {W.}~\bibnamefont {Meevasana}}, \bibinfo {author}
  {\bibfnamefont {N.~J.~C.}\ \bibnamefont {Ingle}}, \bibinfo {author}
  {\bibfnamefont {Y.}~\bibnamefont {Maeno}}, \bibinfo {author} {\bibfnamefont
  {Z.-X.}\ \bibnamefont {Shen}}, \ and\ \bibinfo {author} {\bibfnamefont
  {A.~P.}\ \bibnamefont {Mackenzie}},\ }\href@noop {} {\bibfield  {journal}
  {\bibinfo  {journal} {Phys. Rev. Lett.}\ }\textbf {\bibinfo {volume} {99}},\
  \bibinfo {pages} {187001} (\bibinfo {year} {2007})}\BibitemShut {NoStop}%
\bibitem [{\citenamefont {Kittaka}\ \emph {et~al.}(2010)\citenamefont
  {Kittaka}, \citenamefont {Taniguchi}, \citenamefont {S.Yonezawa},
  \citenamefont {Yaguchi},\ and\ \citenamefont {Maeno}}]{kittaka10}%
  \BibitemOpen
  \bibfield  {author} {\bibinfo {author} {\bibfnamefont {S.}~\bibnamefont
  {Kittaka}}, \bibinfo {author} {\bibfnamefont {H.}~\bibnamefont {Taniguchi}},
  \bibinfo {author} {\bibnamefont {S.Yonezawa}}, \bibinfo {author}
  {\bibfnamefont {H.}~\bibnamefont {Yaguchi}}, \ and\ \bibinfo {author}
  {\bibfnamefont {Y.}~\bibnamefont {Maeno}},\ }\href@noop {} {\bibfield
  {journal} {\bibinfo  {journal} {Phys. Rev. B}\ }\textbf {\bibinfo {volume}
  {81}},\ \bibinfo {pages} {180510(R)} (\bibinfo {year} {2010})}\BibitemShut
  {NoStop}%
\bibitem [{\citenamefont {Hicks}\ \emph {et~al.}(2014)\citenamefont {Hicks},
  \citenamefont {Brodsky}, \citenamefont {Yelland}, \citenamefont {Gibbs},
  \citenamefont {Bruin}, \citenamefont {Barber}, \citenamefont {Edkins},
  \citenamefont {Nishimura}, \citenamefont {Yonezawa}, \citenamefont {Maeno},\
  and\ \citenamefont {Mackenzie}}]{hicks14}%
  \BibitemOpen
  \bibfield  {author} {\bibinfo {author} {\bibfnamefont {C.~W.}\ \bibnamefont
  {Hicks}}, \bibinfo {author} {\bibfnamefont {D.~O.}\ \bibnamefont {Brodsky}},
  \bibinfo {author} {\bibfnamefont {E.~A.}\ \bibnamefont {Yelland}}, \bibinfo
  {author} {\bibfnamefont {A.~S.}\ \bibnamefont {Gibbs}}, \bibinfo {author}
  {\bibfnamefont {J.~A.~N.}\ \bibnamefont {Bruin}}, \bibinfo {author}
  {\bibfnamefont {M.~E.}\ \bibnamefont {Barber}}, \bibinfo {author}
  {\bibfnamefont {S.~D.}\ \bibnamefont {Edkins}}, \bibinfo {author}
  {\bibfnamefont {K.}~\bibnamefont {Nishimura}}, \bibinfo {author}
  {\bibfnamefont {S.}~\bibnamefont {Yonezawa}}, \bibinfo {author}
  {\bibfnamefont {Y.}~\bibnamefont {Maeno}}, \ and\ \bibinfo {author}
  {\bibfnamefont {A.~P.}\ \bibnamefont {Mackenzie}},\ }\href@noop {} {\bibfield
   {journal} {\bibinfo  {journal} {Science}\ }\textbf {\bibinfo {volume}
  {344}},\ \bibinfo {pages} {283} (\bibinfo {year} {2014})}\BibitemShut
  {NoStop}%
\bibitem [{\citenamefont {Taniguchi}\ \emph {et~al.}(2015)\citenamefont
  {Taniguchi}, \citenamefont {Nishimura}, \citenamefont {Goh}, \citenamefont
  {Yonezawa},\ and\ \citenamefont {Maeno}}]{taniguchi15}%
  \BibitemOpen
  \bibfield  {author} {\bibinfo {author} {\bibfnamefont {H.}~\bibnamefont
  {Taniguchi}}, \bibinfo {author} {\bibfnamefont {K.}~\bibnamefont
  {Nishimura}}, \bibinfo {author} {\bibfnamefont {S.~K.}\ \bibnamefont {Goh}},
  \bibinfo {author} {\bibfnamefont {S.}~\bibnamefont {Yonezawa}}, \ and\
  \bibinfo {author} {\bibfnamefont {Y.}~\bibnamefont {Maeno}},\ }\href@noop {}
  {\bibfield  {journal} {\bibinfo  {journal} {J. Phys. Soc. Jpn.}\ }\textbf
  {\bibinfo {volume} {84}},\ \bibinfo {pages} {014707} (\bibinfo {year}
  {2015})}\BibitemShut {NoStop}%
\bibitem [{\citenamefont {Matsumoto}\ and\ \citenamefont
  {Sigrist}(1999)}]{matsumoto99}%
  \BibitemOpen
  \bibfield  {author} {\bibinfo {author} {\bibfnamefont {M.}~\bibnamefont
  {Matsumoto}}\ and\ \bibinfo {author} {\bibfnamefont {M.}~\bibnamefont
  {Sigrist}},\ }\href@noop {} {\bibfield  {journal} {\bibinfo  {journal} {J.
  Phys. Soc. Jpn.}\ }\textbf {\bibinfo {volume} {68}},\ \bibinfo {pages} {994}
  (\bibinfo {year} {1999})},\ \bibinfo {note} {{J. Phys. Soc. Jpn.}
  \textbf{68}, 3120 (1999)}\BibitemShut {NoStop}%
\bibitem [{\citenamefont {Furusaki}\ \emph {et~al.}(2001)\citenamefont
  {Furusaki}, \citenamefont {Matsumoto},\ and\ \citenamefont
  {Sigrist}}]{furusaki01}%
  \BibitemOpen
  \bibfield  {author} {\bibinfo {author} {\bibfnamefont {A.}~\bibnamefont
  {Furusaki}}, \bibinfo {author} {\bibfnamefont {M.}~\bibnamefont {Matsumoto}},
  \ and\ \bibinfo {author} {\bibfnamefont {M.}~\bibnamefont {Sigrist}},\
  }\href@noop {} {\bibfield  {journal} {\bibinfo  {journal} {Phys. Rev. B}\
  }\textbf {\bibinfo {volume} {64}},\ \bibinfo {pages} {054514} (\bibinfo
  {year} {2001})}\BibitemShut {NoStop}%
\bibitem [{\citenamefont {Bouhon}\ and\ \citenamefont
  {Sigrist}(2014)}]{bouhon14}%
  \BibitemOpen
  \bibfield  {author} {\bibinfo {author} {\bibfnamefont {A.}~\bibnamefont
  {Bouhon}}\ and\ \bibinfo {author} {\bibfnamefont {M.}~\bibnamefont
  {Sigrist}},\ }\href@noop {} {\bibfield  {journal} {\bibinfo  {journal} {Phys.
  Rev. B}\ }\textbf {\bibinfo {volume} {90}},\ \bibinfo {pages} {220511(R)}
  (\bibinfo {year} {2014})}\BibitemShut {NoStop}%
\bibitem [{\citenamefont {Huang}\ \emph {et~al.}(2015)\citenamefont {Huang},
  \citenamefont {Lederer}, \citenamefont {Taylor},\ and\ \citenamefont
  {Kallin}}]{huang15}%
  \BibitemOpen
  \bibfield  {author} {\bibinfo {author} {\bibfnamefont {W.}~\bibnamefont
  {Huang}}, \bibinfo {author} {\bibfnamefont {S.}~\bibnamefont {Lederer}},
  \bibinfo {author} {\bibfnamefont {E.}~\bibnamefont {Taylor}}, \ and\ \bibinfo
  {author} {\bibfnamefont {C.}~\bibnamefont {Kallin}},\ }\href@noop {}
  {\bibfield  {journal} {\bibinfo  {journal} {Phys. Rev. B}\ }\textbf {\bibinfo
  {volume} {91}},\ \bibinfo {pages} {094507} (\bibinfo {year}
  {2015})}\BibitemShut {NoStop}%
\bibitem [{\citenamefont {Raghu}\ \emph {et~al.}(2010)\citenamefont {Raghu},
  \citenamefont {Kapitulnik},\ and\ \citenamefont {Kivelson}}]{raghu10}%
  \BibitemOpen
  \bibfield  {author} {\bibinfo {author} {\bibfnamefont {S.}~\bibnamefont
  {Raghu}}, \bibinfo {author} {\bibfnamefont {A.}~\bibnamefont {Kapitulnik}}, \
  and\ \bibinfo {author} {\bibfnamefont {S.~A.}\ \bibnamefont {Kivelson}},\
  }\href@noop {} {\bibfield  {journal} {\bibinfo  {journal} {Phys. Rev. Lett.}\
  }\textbf {\bibinfo {volume} {105}},\ \bibinfo {pages} {136401} (\bibinfo
  {year} {2010})}\BibitemShut {NoStop}%
\bibitem [{\citenamefont {Imai}\ \emph {et~al.}(2012)\citenamefont {Imai},
  \citenamefont {Wakabayashi},\ and\ \citenamefont {Sigrist}}]{imai12}%
  \BibitemOpen
  \bibfield  {author} {\bibinfo {author} {\bibfnamefont {Y.}~\bibnamefont
  {Imai}}, \bibinfo {author} {\bibfnamefont {K.}~\bibnamefont {Wakabayashi}}, \
  and\ \bibinfo {author} {\bibfnamefont {M.}~\bibnamefont {Sigrist}},\
  }\href@noop {} {\bibfield  {journal} {\bibinfo  {journal} {Phys. Rev. B}\
  }\textbf {\bibinfo {volume} {85}},\ \bibinfo {pages} {174532} (\bibinfo
  {year} {2012})}\BibitemShut {NoStop}%
\bibitem [{\citenamefont {Qin}\ \emph {et~al.}(2011)\citenamefont {Qin},
  \citenamefont {Niu},\ and\ \citenamefont {Shi}}]{qin11}%
  \BibitemOpen
  \bibfield  {author} {\bibinfo {author} {\bibfnamefont {T.}~\bibnamefont
  {Qin}}, \bibinfo {author} {\bibfnamefont {Q.}~\bibnamefont {Niu}}, \ and\
  \bibinfo {author} {\bibfnamefont {J.}~\bibnamefont {Shi}},\ }\href@noop {}
  {\bibfield  {journal} {\bibinfo  {journal} {Phys. Rev. Lett.}\ }\textbf
  {\bibinfo {volume} {107}},\ \bibinfo {pages} {236601} (\bibinfo {year}
  {2011})}\BibitemShut {NoStop}%
\bibitem [{\citenamefont {Sumiyoshi}\ and\ \citenamefont
  {Fujimoto}(2013)}]{sumiyoshi13}%
  \BibitemOpen
  \bibfield  {author} {\bibinfo {author} {\bibfnamefont {H.}~\bibnamefont
  {Sumiyoshi}}\ and\ \bibinfo {author} {\bibfnamefont {S.}~\bibnamefont
  {Fujimoto}},\ }\href@noop {} {\bibfield  {journal} {\bibinfo  {journal} {J.
  Phys. Soc. Jpn.}\ }\textbf {\bibinfo {volume} {82}},\ \bibinfo {pages}
  {023602} (\bibinfo {year} {2013})}\BibitemShut {NoStop}%
\end{thebibliography}%
\end{document}